\documentclass[pra,twocolumn,showpacs,floatfix,superscriptaddress,aps]{revtex4}
\usepackage{amsmath}
\usepackage{makeidx}
\usepackage{amssymb}
\usepackage{graphicx}

\usepackage[usenames]{color}
\usepackage[normalem]{ulem}

\usepackage{hyperref}
\usepackage[T1]{fontenc} 
\usepackage{soul}
\begin{document}
%%%%%%%%%%%%%%%%%%%%%%%%%%%%%%%%%%%%%%%%%%%%%%%%%%%%%%%%%%%%%%%%%%%%%%%%%%%%%%
%%%%                     Title and authors                                %%%%
%%%%%%%%%%%%%%%%%%%%%%%%%%%%%%%%%%%%%%%%%%%%%%%%%%%%%%%%%%%%%%%%%%%%%%%%%%%%%%

\title{Analytic models for density of a  ground-state spinor condensate}

\author{Sandeep Gautam\footnote{sandeepgautam24@gmail.com}}
\author{S. K. Adhikari\footnote{adhikari44@yahoo.com, 
        URL  http://www.ift.unesp.br/users/adhikari}}
\affiliation{Instituto de F\'{\i}sica Te\'orica, Universidade Estadual
             Paulista - UNESP, \\ 01.140-070 S\~ao Paulo, S\~ao Paulo, Brazil}
      
%%%%%%%%%%%%%%%%%%%%%%%%%%%%%%%%%%%%%%%%%%%%%%%%%%%%%%%%%%%%%%%%%%%%%%%%%%%%%%
%%%%%%%%%%                    Abstract                             %%%%%%%%%%%
%%%%%%%%%%%%%%%%%%%%%%%%%%%%%%%%%%%%%%%%%%%%%%%%%%%%%%%%%%%%%%%%%%%%%%%%%%%%%%

\date{\today}
\begin{abstract} 
We demonstrate that the ground state of a trapped spin-1 and spin-2 spinor 
ferromagnetic Bose-Einstein condensate (BEC) can be well approximated  by a 
single decoupled Gross-Pitaevskii (GP) equation. Useful analytic models for the 
ground-state densities of    
ferromagnetic BECs are obtained from the Thomas-Fermi approximation (TFA) to this 
decoupled equation. Similarly, for the  ground states of  spin-1 anti-ferromagnetic and spin-2 anti-ferromagnetic and 
cyclic BECs, some of the spin component densities are zero which reduces the coupled 
GP equation to a simple reduced form.
Analytic models for ground state densities are also obtained 
for   anti-ferromagnetic and 
cyclic BECs from the TFA to the respective reduced GP equations. The analytic densities    are illustrated and compared with the full numerical solution of the GP 
equation with realistic experimental parameters.
\end{abstract}

\maketitle

%%%%%%%%%%%%%%%%%%%%%%%%%%%%%%%%%%%%%%%%%%%%%%%%%%%%%%%%%%%%%%%%%%%%%%%%%%%%%%%
%%%%%                        Introduction                             %%%%%%%%%
%%%%%%%%%%%%%%%%%%%%%%%%%%%%%%%%%%%%%%%%%%%%%%%%%%%%%%%%%%%%%%%%%%%%%%%%%%%%%%%
\section{Introduction}
The advent of the optical traps paved the way for the first realization of a
Bose-Einstein condensate (BEC) with internal spin degrees of freedom 
\cite{Stamper-Kurn1}, also known as a spinor BEC. Since then, a lot of 
theoretical and experimental studies 
have been performed  on the spinor BECs \cite{Kawaguchi,ueda,Stamper-Kurn2}. 
In contrast to a 
scalar BEC, which is characterized by a single interaction 
parameter, the spin-$1$ \cite {Ohmi} and spin-$2$ \cite{Ciobanu,Ueda}   BECs have, 
respectively, two and three interaction parameters. 
Depending on the relative strength of the interaction parameters, a 
spin-1 BEC {in the absence of external magnetic field} can be either 
in a {\em ferromagnetic} or an {\em anti-ferromagnetic} phase \cite{Ohmi}.
In the presence of magnetic field the ground state phase diagram of spin-1
condensate has been investigated both for uniform \cite{Stenger,Romano,Zhang} 
and trapped systems \cite{Zhang,Matuszewski}.  
Similarly, in the absence of external magnetic field, a spin-2 BECs 
can be in one of the possible three ground state phases: 
{\em ferromagnetic}, {\em anti-ferromagnetic} and {\em cyclic} \cite{Ciobanu,Ueda}. 
The spin-1 and spin-2 BECs are described by three- and five-component complex order parameters, respectively, thus leading to coupled mean-field Gross-Pitaevskii (GP) equations  
involving three- and five-component  wave functions, which{, unlike in a scalar BEC,} could be complex in general. 
{A numerical solution of these equations 
could be cumbersome  for both spin-1 \cite{Wang-1,sadhan} and spin-2 \cite{Wang-2,GA-2} 
BECs.} In this paper, we propose simple and useful analytic models for the densities of 
ground states of 
quasi-one-dimensional (quasi-1D), circularly-symmetric quasi-two-dimensional (quasi-2D) \cite{luca}, and spherically-symmetric three-dimensional (3D) spin-1 and spin-2 spinor BECs.
{ Here, we consider  nearly-overlapping spatially-symmetric ground states only. Phase-separated spatially-asymmetric profiles \cite{sadhan} do not appear as ground states and will not be considered. }

The two interaction parameters for a spin-1 BEC are 
$c_0\propto (a_0 + 2a_2)/3$ and $c_1 \propto (a_2 - a_0)/3$ \cite{Ohmi}, whereas
the three interaction parameters for a spin-2 BEC are 
$c_0\propto (4a_2+3a_4)/7$, $c_1 \propto (a_4-a_2)/7$, and
$c_2 \propto (7a_0-10a_2+3a_4)/7$ \cite{Ciobanu,Ueda}, where $a_0,a_2$, and $a_4$ are $s$-wave scattering
lengths in total spin $f_{\rm tot} = 0,2$, and $4$ channels, respectively.
For a ferromagnetic BEC, e.g. for $c_1<0$  for a spin-1 BEC, and for $c_1<0$ and $c_2>20c_1$ 
for a spin-2 BEC, we find that to a very good approximation the densities for different 
spin components {$m_f$} of the ground-state wave function  with  magnetization $m$
are multiples of each other. 
This  allows one to replace the coupled GP equation
for the ground-state wave function 
  by a single  partial differential equation, which we call
the decoupled-mode (DM) equation.
 On the other hand, for an anti-ferromagnetic
BEC, e.g. for $c_1>0$  for a spin-1 BEC, and  for $c_2<0$   and $c_2<20c_1$ 
 for a spin-2 BEC, we find that the densities, for some of the spin components,
    of the ground-state wave function  with  magnetization $m$ are identically zero,
thus reducing the original coupled GP equation to a system of  
 two coupled equations for
any non-zero magnetization. Similarly, for a cyclic BEC, e.g.  for $c_1>0$ and $c_2>0$
for a spin-2 BEC, 
the five-component GP equation  reduces to a system of two or three coupled  equations. 
These reduced GP equations and the DM equation 
for the ground state of a spinor BEC in different parameter domains, valid in all spatial 
dimensions,  are solved in the Thomas-Fermi approximation (TFA) (or local-density approximation)
to yield simple  analytic models for the ground-state densities of   spin-$1$ and spin-2 spinor BECs.

{ 
The TFA is applicable when the interaction energy in the GP equation is much larger than the 
kinetic energy term, so that the latter   could be neglected, thus leading to simple 
analytic formulae for the condensate densities   \cite{Baym}. 
In a repulsive scalar BEC, applicability of TFA requires that the size of the condensate $R$ 
is much larger than the oscillator length $l_0$, i.e.
 $R/l_0>>1$  \cite{Dalfovo, Pethick}. The spatial extent of the BEC in units of $l_0$ is
\begin{equation}\label{tfasize}
l_{{\cal D}} = \frac{R}{l_0}\sim \left(\frac{Na}{l_0}\right)^{1/({\cal D}+2)},
\end{equation} 
where ${\cal D}=1, 2, 3$ is the dimensionality of the space \cite{Dalfovo, Pethick}.
The criterion is satisfied if the dimensionless
parameter $Na/l_0>>1$. The ratio $Na/l_0$ is a measure of the strength of repulsive
interaction. For a spinor BEC, the applicability of TFA for $m_f$ component
requires that its spatial extent is much larger than $l_0$.
%In this limit, the ratio of interaction to
%kinetic energy is
% \begin{equation}\label{cond}
%\frac{E_{\mathrm{kin}}}{E_{\mathrm{int}}} \sim8\pi N \frac{a}{l_0}.
% \end{equation}
%An estimate for the validity of the TFA for repulsive atomic interaction
%can be obtained using the Heisenberg uncertainty relation $\Delta p\sim \hbar/\Delta x$. 
%For a trapped BEC, $\Delta x\sim R$ and kinetic energy $E_{\mathrm{kin}}\sim 
%N\hbar^2/2{M}R^2$
%{for a scalar BEC,} 
%where {$M$} is the mass, $N$ the number of atoms,  $l_0$ the harmonic oscillator length, and $R$ is the spatial extention of the BEC \cite{kinetic}. 
%In the mean-field approach the interaction for sufficiently large condensate kinetic
%energy can be negelected, and the equilibrium spatial extent of the cloud can be obtained by minimizing the sum total of interaction 
%energy $E_{\mathrm{int}}= N U_0 \rho$ and potential energy where $U_0=4\pi \hbar^2 a/M,$  
%$\rho \sim N/R^3$ is the density
%and $a$ the {$s$-wave} 
%scattering length. For a sufficiently large $R$, kinetic energy is negligible in comparison to the interaction and potential energies. The spatial extent of the 
%ground state can be obtained by minimizing the sum total of interaction and potential energies 

%Hence we have
%Consequently, the kinetic energy can be neglected in comparison to the interaction 
%energy to validate the TFA, when $Na>>l_0$. {The ratio $Na/l_0$ is the dimensionless
%measure of the strength of interaction.} 
There have been few studies to include the neglected 
kinetic energy contribution in the TFA \cite{Dalfovo,kinetic}.
Earlier, the  TFA was used to study the ground state 
properties of  binary condensates \cite{Ho} {and spin dynamics in quasi-1D spin-1 condensate \cite{Mur-Petit}.} 
Spin-orbit-coupled pseudospin-$1/2$ BECs under rotation have also been theoretically investigated 
using the TFA \cite{Aftalion}. 
}

We use the experimentally realizable trapping potential and interaction parameters
 to illustrate the present analytic models for  ground-state densities in different parameter domains. 
In the case of a spin-1
BEC, 
the background scattering lengths of $^{87}$Rb and $^{23}$Na fall in the
ferromagnetic \cite{kempen,Chang} and anti-ferromagnetic \cite{Black} domains, 
respectively, and  we use these to study the ground state properties. 
In the case of a spin-$2$ BEC, we employ {$^{23}$Na and $^{83}$Rb BECs} for the illustration.
{The background scattering lengths of spin-2 $^{23}$Na and $^{83}$Rb correspond to the 
anti-ferromagnetic and ferromagnetic phases, respectively \cite{Ciobanu}.} 
By tuning one of the scattering lengths of $^{23}$Na, one can move 
from anti-ferromagnetic to either ferromagnetic or cyclic phase. Experimentally, such a
change of single scattering length can be achieved by exploiting 
magnetic \cite{Inouye} and optical \cite{Chin} Feshbach resonance techniques. The results of the analytic
models are also validated by a numerical solution of the original mean-field GP equations for quasi-1D and quasi-2D traps.    

  In Sec. \ref{Sec-II}, we present  the
full mean-field  GP equations for spin-1 and spin-2 BECs  and derive the 
reduced mean-field GP equations for the ground-state wave function in the different 
parameter domains. By assuming that 
the component wavefunctions of a ferromagnetic BEC are proportional to each other, 
which is indeed
the case as suggested by numerical simulations, we derive the DM equation. 
By minimizing the $c_1$- and $c_2$-dependent energy
terms for the ground-state wave function, we obtain the reduced GP equations in all 
parameter domains.  In Secs. \ref{Sec-III} and  \ref{Sec-IV}
we obtain the  analytic models for  spin-1 and spin-2 ground-state BECs, respectively, by  employing the 
TFA to the reduced GP and the DM equations. A comparison of the analytic densities with the numerical densities 
obtained from the full GP equations leads to a very satisfactory agreement.
In Sec. \ref{V}, we present a summary and concluding remarks. Some of the technical details about 
the derivation of the   DM equation and the reduced GP equations in different parameter domains are presented in  Appendix A and B.

  %%%%%%%%%%%%%%%%%%%%%%%%%%%%%%%%%%%%%%%%%%%%%%%%%%%%%%%%%%%%%%%%%%%%%%%%%%%%%%%%%%%%

\section{Reduced Mean-Field  Equations}
\label{Sec-II}
\subsection{Spin-1 BEC}
  The coupled GP equations for different spin components $m_f=\pm 1,0$,
for a spin-1 BEC of $N$ atoms of mass $M$ each
can be written in dimensionless form as \cite{Kawaguchi}
\begin{align}
\mu_{\pm 1} \phi_{\pm 1}(\mathbf x) &=
 {\cal H}\phi_{\pm 1}(\mathbf x) +c^{}_0\rho\phi_{\pm 1}(\mathbf x)
\pm   c^{}_1F_z\phi_{\pm 1}(\mathbf x) \nonumber \\
&+ ({c^{}_1}/{\sqrt{2}}) F_{\mp}\phi_0(\mathbf x),
 \label{gps-1}\\
\mu_0 \phi_0(\mathbf x) &=
{\cal H}\phi_0(\mathbf x)    +c^{}_0\rho\phi_{0}(\mathbf x)
+ (c_1/{\sqrt 2}) [F_{-}\phi_{-1}(\mathbf x)\nonumber \\&+F_{+}\phi_{+1}(\mathbf x)]\label{gps-2}, 
\end{align}
where{
\begin{align}
F_{\pm}\equiv & F_x\pm F_y=\sqrt 2[\phi_{\pm 1}^*(\mathbf x)\phi_0(\mathbf x)+\phi_0^*(\mathbf x)\phi_{\mp 1}(\mathbf x)]\label{fpmspin1}, \\
  F_z=&\rho_{+1}(\mathbf x)-\rho_{-1}(\mathbf x)\label{fzspin1},
\quad
{\cal H}=\biggr[ -\frac{1}{2}\nabla^2
 +{V}({\mathbf x})\biggr],
\end{align}
an}d  the component density $
\rho_j=|\phi_j(\mathbf x)|^2$ with $j=\pm 1, 0$, the total density  $\rho=\sum_{j}\rho_j,$ and $\mu_{\pm 1},\mu_0$
are the respective chemical potentials and $^*$ denotes complex conjugate.
In 3D, the interaction parameters, Laplacian, and trapping potential are defined
 as
\begin{align}
c_0 =& \frac{4\pi N (a_0+2 a_2)}{3 l_0},~
c_1 = \frac{4\pi N (a_2-a_0)}{3 l_0},\\
\nabla^2=&\frac{\partial ^2}{\partial  x^2} + \frac{\partial ^2}{\partial y^2}+\frac{\partial ^2}{\partial z^2},
V(\mathbf x)=\frac{x^2+\beta^2y^2+\gamma^2z^2}{2},
\end{align} 
with ${\bf x}\equiv\{{x,y,z}\}$.
{   Here}
$l_0=\sqrt{\hbar/(M\omega_x)}$, $\beta = \omega_y/\omega_x$, $\gamma = \omega_z/\omega_x$, where $\omega_x,\omega_y,\omega_z$
are the confining trap frequencies in $x,y,z$ directions, respectively. 
 When the trapping frequency along
one axis, say $\omega_z$, is much larger than the geometric mean of other two, i.e., 
$\omega_z\gg\sqrt{\omega_x \omega_y}$, then one can approximate the 
Eqs. (\ref{gps-1}) -(\ref{gps-2}) by quasi two-dimensional (2D) equations which can be obtained
by substituting \cite{luca}
\begin{align}
c_0 =& \frac{2N \sqrt{2\pi}(a_0+2 a_2)}{3 l_z},~
c_1 = \frac{2N \sqrt{2\pi}({a_2-a_0})}{3 l_z},\\
\nabla^2=&\frac{\partial ^2}{\partial x^2} + \frac{\partial ^2}{\partial y^2},~
V(\mathbf x)=\frac{x^2+\beta^2y^2}{2},~
{\mathbf x }\equiv\{x,y\},\label{nabla_q2d}
\end{align}
in Eqs. (\ref{gps-1}) -(\ref{gps-2}), here $l_z = \sqrt{\hbar/(M\omega_z)}$.
Similarly, if the trapping frequencies along two axes, say $y$ and $z$, are much larger than 
the third frequency $\omega_x$, Eqs. (\ref{gps-1}) -(\ref{gps-2}) can be approximated by quasi-1D
equations which can be obtained by substituting
\begin{align}
c_0 &= \frac{2 N (a_0+2 a_2)l_0}{3 l_{yz}^2},~
c_1 = \frac{2N (a_2-a_0)l_0}{3 l_{yz}^2},\\
\nabla^2&=\frac{\partial ^2}{\partial x^2},~
V(\mathbf x)=\frac{x^2}{2},~
{\mathbf x }\equiv x,\label{nabla_q1d}
\end{align}
where $l_{yz}=\sqrt{\hbar/(M\omega_{yz})}$ and $\omega_{yz}= \sqrt{\omega_y\omega_z}$.
Here length is measured in units of $l_0$, density in units of $l_0^{-\cal D}$
and chemical potential in units of $\hbar\omega_x$, where ${\cal D}=1,2,3$ is the dimensionality of space.
The total density is normalized to unity $\int d{\mathbf x} \rho(\mathbf x)=1$.
The volume element $d{\bf x}= 2 dX$ in 1D, $2\pi X dX$ in 
2D with circular symmetry, and $4 \pi X^2 dX$ in 3D with spherical symmetry,
where $X=|{\bf x}|$ is the length of the vector $\bf x.$
In this paper, we consider isotropic 3D and isotropic quasi-2D traps, i.e., $\beta = \gamma = 1$
for 3D traps and $\beta = 1\ll\gamma$ for quasi-2D traps.

%The  quasi-1D  coupled GP equations for different spin components $m_f=\pm 1,0$,
%for a spin-1 BEC of $N$ atoms of mass $M$ each
%in the presence of a strong transverse trap can be written  in dimensionless form as:
%\begin{align}
%\mu_{\pm 1} \phi_{\pm 1}(x) &=
% {\cal H}\phi_{\pm 1}(x) +c_0\rho\phi_{\pm 1}(x)
%\pm   c_1F_z\phi_{\pm 1}(x) \nonumber \\
%&+ ({c_1}/{\sqrt{2}}) F_{\mp}\phi_0(x),
% \label{gps-1}\\
%\mu_0 \phi_0(x) &=
%{\cal H}\phi_0(x)    +c_0\rho\phi_{0}(x)
%+ ({{c}_1}/{\sqrt 2}) [F_{-}\phi_{-1}(x)\nonumber \\&+F_{+}\phi_{+1}(x)]\label{gps-2}, 
%\end{align}
%where 
%\begin{align}
%F_{\pm}\equiv & F_x\pm F_y=\sqrt 2[\phi_{\pm 1}^*(x)\phi_0(x)+\phi_0^*(x)\phi_{\mp 1}(x)]\label{fpmspin1}, \\
%  F_z=&|\phi_{+1}(x)|^2-|\phi_{-1}(x)|^2\label{fzspin1},\\
%{\cal H}=&\biggr[ -\frac{1}{2}\frac{\partial^2}{\partial {x}^2}
% +{V}({x})\biggr],
%\end{align}
%and the trap $V(x)=x^2/2,$ the interaction parameters $c_0= 2 N (a_0+2 a_2)l_0/(3 l_{yz}^2), c_1= 2N 
%(a_2-a_0)l_0/(3 l_{yz}^2),$ the component density $
%\rho_j=|\phi_j(x)|^2$ with $j=\pm 1, 0$, the total density  $\rho=\sum_{j}\rho_j,$ and $\mu_{\pm 1},\mu_0$ 
%are the respective chemical potentials and $^*$ denotes complex conjugate. Here  
%$l_0=\sqrt{\hbar/(M\omega_x)},
%l_{yz}=\sqrt{\hbar/(M\omega_{yz})}, \omega_{yz}= \sqrt{\omega_y\omega_z}$, where $\omega_x,\omega_y,\omega_z$ 
%are the confining trap frequencies in $x,y,z$ directions, respectively, length is measured in units of $l_0$,
%and chemical potential in units of $\hbar\omega_x$.
%The total density is normalized to unity $\int dx \rho(x)=1$.

Numerical calculation for the ground-state densities of  a 
 of  a
ferromagnetic BEC
($c_1<0$) has revealed that the component densities are essentially {multiples} of each other. This opens the possibility of 
writing a single decoupled mode (DM) equation for the  wave-function
$\phi_{\mathrm{DM}}$ 
for the ferromagnetic BEC and obtain the component  wave functions  as multiples of this wave function according to   
\begin{eqnarray}\label{sma}
 \phi_{j}(\mathbf x) = \alpha _{j}\phi_{\rm DM}(\mathbf x), \quad j=\pm1,0,
\end{eqnarray}
{where $\alpha_j$'s, in general, are complex numbers.}
   The conditions (\ref{sma})
when substituted in Eqs. (\ref{gps-1}) and (\ref{gps-2}) lead to three different equations for the 
same wave function $\phi_{\mathrm{DM}}$. A consistency requirement on these three equations leads to the  single   decoupled-mode   (DM) equation for the wave function $\phi_{\mathrm{DM}}$:
\begin{eqnarray}\label{DM-1}
 \mu \phi_{\mathrm{DM}}(\mathbf x)= \biggr[ -\frac{1}{2}\nabla^2%\frac{\partial^2}{\partial {x}^2}
 +{V}({\mathbf x})+C{\phi^2_{\mathrm{DM}}(\mathbf x)}\biggr]\phi_{\mathrm{DM}}(\mathbf x),
\end{eqnarray}
with $C\equiv {\cal C}_{I}=c_0+c_1$ and normalization $\int \rho_{\mathrm{DM}}({\bf x}) d{\bf x}=1,$
provided that 
\begin{eqnarray}\label{dist-1}
|\alpha_{\pm 1}| &=&\frac{1\pm m}{2},\quad  |\alpha_0|=\frac{\sqrt{1-m^2}}{\sqrt{2}},\\
m & \equiv& \int d{\bf x} [\rho_{+1}-\rho_{-1}  ]=|\alpha_{+1}|^2 - |\alpha_{-1}|^2,
\end{eqnarray}
where  $m$ is the magnetization. 
 % here $l$ denotes
%the distance of the edge of the condensate from the trap center.}
An equation  similar to  Eq. (\ref{DM-1}) with $C=c_0\sim (a_0+2a_2)/3$, known as the   single-mode approximation (SMA) \cite{Law},
was obtained before as an approximation to Eqs. 
(\ref{gps-1})-(\ref{gps-2}).  The component densities were then obtained 
using Eq. (\ref{dist-1}). In the DM model we have a different $C\equiv {\cal C}_I=(c_0+c_1)\sim a_2$, which 
is independent of $a_0$.  
Equation (\ref{DM-1}) was previously obtained by Yi {\it et al.} \cite{Yi}  
as an improvement over the SMA.
The breakdown of the 
single-mode approximation  for trapped spin-1 condensates in the presence of magnetic 
field has also been theoretically investigated \cite{Zhang}.

Provided that {\em ansatz} (\ref{sma}) holds,
distribution (\ref{dist-1}) can  be obtained independently from 
a consideration of $c_1$-dependent  energy minimization 
for a ferromagnetic BEC  as shown in  Appendix A.
The DM equation is very useful for finding the ground state of a ferromagnetic BEC  where
all density components are non-zero and this procedure can also be readily generalized to higher-spin cases as shown in Sec. \ref{IB} for a spin-2 ferromagnetic BEC. 
 
For the ground-state of 
an anti-ferromagnetic BEC ($c_1>0$) {with non-zero magnetization}, 
energy minimization requires that $\phi_0({\mathbf x})=0$, viz. Appendix A.
Then     the  normalization and 
   magnetization conditions 
yield 
\begin{eqnarray}\label{norm-1}
\int \rho_{\pm 1} d{\bf x} = \frac{ 1\pm m}{2}, \quad \rho_0=0.
\end{eqnarray}
{ For $m=0$, besides the aforementioned state, there is another degenerate 
state where all the atoms are in $m_f = 0$ component, i.e. $\rho_{\pm 1}=0$ and $\int \rho_0 d{\bf x} = 1$.
}
Unlike in a ferromagnetic BEC,
  {\em ansatz} (\ref{sma}) does not hold for an anti-ferromagnetic  BEC for a non-zero magnetization $m$
where different components occupy different spatial extensions. {On the other hand, 
for $m=0$, SMA becomes exact in this phase \cite{Yi}, as the $c_1$-dependent term vanishes.} 
%However, if we assume ansatz (\ref{sma}), a
%$c_1$-dependent energy minimization for an anti-ferromagnetic BEC leads to $\phi_0%(x)=0.$

We will derive the analytic model for a ferromagnetic BEC using  the TFA to the  DM equation (\ref{DM-1}), whereas for an 
anti-ferromagnetic BEC we rely on the TFA to the
GP equation (\ref{gps-1}) with $\phi_0(\mathbf x)=0$ for the same.

\subsection{Spin-2 BEC}

\label{IB}
{The  dimensionless coupled GP equations for different spin components
$m_f=\pm 2,\pm 1,0$,
for a spin-2
BEC can be written as \cite{Kawaguchi}
 \begin{align} 
 \mu_{\pm 2}& \phi_{\pm 2}(\mathbf x) =
{\cal H}\phi_{\pm 2}(\mathbf x) +c_0\rho\phi_{\pm 2}(\mathbf x)+({{c}_2}/{\sqrt{5}}){\Theta}\phi_{\mp 2}^*(\mathbf x)
\nonumber\\
&   +{c}_1\big[{F}_{\mp} \phi_{\pm 1}(\mathbf x)\pm 2{F}_{{z}}\phi_{\pm 2}(\mathbf x)\big] 
  \label{gp_s1}
 ,
\\
 \mu_{\pm 1}& \phi_{\pm 1}(\mathbf x) = 
{\cal H} \phi_{\pm 1} (\mathbf x) +c_0\rho\phi_{\pm 1}(\mathbf x)  -({{c}_2}/{\sqrt{5}}){\Theta}\phi_{\mp 1}^*(\mathbf x)\nonumber\\
 &+{c}_1\big[\sqrt{3/2} {F}_{\mp}\phi_0(\mathbf x)+{F}_{\pm}\phi_{\pm 2} (\mathbf x)
\pm {F}_{{z}}\phi_{\pm 1}(\mathbf x)\big] ,
 \label{gp_s2}
 \\
 \mu_{0}& \phi_{0}(\mathbf x) =
{\cal H}\phi_0(\mathbf x) +c_0\rho\phi_{0}(\mathbf x)   +({{c}_2}/{\sqrt{5}}){\Theta}\phi_{0}^*(\mathbf x)\nonumber\\
  &+c_1\sqrt{3/2}\big[{F}_{-}
 \phi_{-1}(\mathbf x)+{F}_{+}\phi_{+1}(\mathbf x)\big]
  ,\label{gp_s3}
\end{align}
where
\begin{align}
 {F}_{+} =&  {F}_{-}^*= 2(\phi_{+2}^*\phi_{+1}+\phi_{-1}^*\phi_{-2}) \nonumber \\
&+\sqrt{6}(\phi_{+1}^*\phi_0 +\phi_0^*\phi_{-1}),  \\
{F}_{{z}} =& 2(|\phi_{+2}|^2-|\phi_{-2}|^2) + |\phi_{+1}|^2-|\phi_{-1}|^2,  \\
{\Theta} =& \frac{2\phi_{+2}\phi_{-2}-2\phi_{+1}\phi_{-1}+\phi_0^2}{\sqrt{5}}.\label{theta}
\end{align}
Here the interaction parameters %$ c_0= 2N(4a_2+3a_4)l_0/(7l_{yz}^2), c_1=2N(a_4-a_2)l_0/(7l_{yz}^2),
%c_2 = 2N(7a_0-10a_2+3a_4)l_0/(7l_{yz}^2), $
$ c_0= 4\pi N(4a_2+3a_4)/(7l_0), c_1=4\pi N(a_4-a_2)/(7l_0),
c_2 = 4\pi N(7a_0-10a_2+3a_4)/(7l_0), $
$\mu_{\pm 2}, \mu_{\pm 1},$ and $\mu_0$ are the respective chemical potentials.
All repeated variables have the same meaning as 
 in the spin-1 case.
The total density $\rho$ is again normalized to unity. As in the  spin-1 case, 
GP equations in quasi-2D traps can be obtained by using Eqs.~(\ref{nabla_q2d}) and substituting 
$c_0 = 2 N \sqrt{2\pi}(4a_2+3a_4)/(7l_z), c_1=2 N\sqrt{2\pi}(a_4-a_2)/(7l_z),
c_2 = 2 N\sqrt{2\pi}(7a_0-10a_2+3a_4)/(7l_z)$ in Eqs.~(\ref{gp_s1})-(\ref{gp_s3}). 
Similarly, GP equations in {quasi-1D} traps can be obtained by using Eqs.~(\ref{nabla_q1d}) and substituting 
$ c_0= 2N(4a_2+3a_4)l_0/(7l_{yz}^2), c_1=2N(a_4-a_2)l_0/(7l_{yz}^2),
c_2 = 2N(7a_0-10a_2+3a_4)l_0/(7l_{yz}^2)$ in Eqs. (\ref{gp_s1})-(\ref{gp_s3}). 
}

%The  dimensionless quasi-1D coupled GP equations for different spin components 
%$m_f=\pm 2,\pm 1,0$,
%for a spin-2 
%BEC can be written as 
% \begin{align} 
% \mu_{\pm 2}& \phi_{\pm 2}(x) =
%{\cal H}\phi_{\pm 2}(x) +c_0\rho\phi_{\pm 2}(x)+({{c}_2}/{\sqrt{5}}){\Theta}\phi_{\mp 2}^*(x)
%\nonumber\\
%&   +{c}_1\big[{F}_{\mp} \phi_{\pm 1}(x)\pm 2{F}_{{z}}\phi_{\pm 2}(x)\big] 
%  \label{gp_s1}
% ,
%\\
% \mu_{\pm 1}& \phi_{\pm 1}(x) = 
%{\cal H} \phi_{\pm 1} (x) +c_0\rho\phi_{\pm 1}(x)  -({{c}_2}/{\sqrt{5}}){\Theta}\phi_{\mp 1}^*(x)\nonumber\\
% &+{c}_1\big[\sqrt{3/2} {F}_{\mp}\phi_0(x)+{F}_{\pm}\phi_{\pm 2} (x)
%\pm {F}_{{z}}\phi_{\pm 1}(x)\big] ,
% \label{gp_s2}
% \\
% \mu_{0}& \phi_{0}(x) =
%{\cal H}\phi_0(x) +c_0\rho\phi_{0}(x)   +({{c}_2}/{\sqrt{5}}){\Theta}\phi_{0}^*(x)\nonumber\\
%  &+c_1\sqrt{3/2}\big[{F}_{-}
% \phi_{-1}(x)+{F}_{+}\phi_{+1}(x)\big]
%  ,\label{gp_s3}
%\end{align}
%where
%\begin{align}
% {F}_{+} =&  {F}_{-}^*= 2(\phi_{+2}^*\phi_{+1}+\phi_{-1}^*\phi_{-2}) \nonumber \\
%&+\sqrt{6}(\phi_{+1}^*\phi_0 +\phi_0^*\phi_{-1}),  \\
%{F}_{{z}} =& 2(|\phi_{+2}|^2-|\phi_{-2}|^2) + |\phi_{+1}|^2-|\phi_{-1}|^2,  \\
%{\Theta} =& \frac{2\phi_{+2}\phi_{-2}-2\phi_{+1}\phi_{-1}+\phi_0^2}{\sqrt{5}}.\label{theta}
%\end{align}
%Here the interaction parameters $ c_0= 2N(4a_2+3a_4)l_0/(7l_{yz}^2), c_1=2N(a_4-a_2)l_0/(7l_{yz}^2),
%c_2 = 2N(7a_0-10a_2+3a_4)l_0/(7l_{yz}^2), $  
%$\mu_{\pm 2}, \mu_{\pm 1},$ and $\mu_0$ are the respective chemical potentials, $N,{\cal H}, l_0, \rho$ and 
%$l_{yz}$ have the same meaning  as in the spin-1 case.  
%The total density $\rho$ is again normalized to unity. 

In the DM, for a ferromagnetic BEC ($c_1<0, c_2>20c_1$) with all non-zero component 
densities, 
if we   {substitute} the {\em ansatz}
\begin{align}\label{comp-2}
\phi_{j}= \alpha_{j}\phi_{\mathrm{DM}},\quad j=\pm2,\pm1,0,  
\end{align}
in Eqs. (\ref{gp_s1}), (\ref{gp_s2}), and (\ref{gp_s3}), we obtain five independent equations for $\phi_{\mathrm{DM}}$.
A consistency requirement among these five equations for the $c_1$-dependent terms 
leads to the  DM  equation  (\ref{DM-1}) with $C\equiv {\cal C}_{II}=(c_0+4c_1)$,
 provided that 
\begin{align}\label{comp-3}
|\alpha _{\pm 2}| &= \frac{(2\pm m)^2}{16 },
\\
|\alpha _{\pm 1}| &= \frac{ \sqrt{4-m^2} (2\pm m)  }{8 },
\\
|\alpha_0| &= \frac{1}{8} \sqrt{\frac{3}{2}} \left(4-m^2\right),\label{comp-4}
\end{align}
with magnetization and normalization conditions
\begin{eqnarray}\label{mag}
m&\equiv & \int d{\bf x} [ 2(\rho_{+2}-\rho_{-2})+ (\rho_{+1}-\rho_{-1})],\\
&=& 2(|\alpha _{+ 2}|^2-|\alpha _{- 2}|^2)+ (|\alpha _{+ 1}|^2-|\alpha _{- 1}|^2),\\
1&=& |\alpha _{+ 2}|^2+|\alpha _{- 2}|^2+ |\alpha _{+ 1}|^2+|\alpha _{- 1}|^2+|\alpha_0|^2.\label{nor}
\end{eqnarray}
In the DM model for a spin-2 BEC $C\equiv {\cal C}_{II}\sim a_4$ is independent of the scattering lengths $a_0$ and $a_2$, with $a_4$ playing the role of scattering 
length in an equivalent scalar BEC.  
With the coefficients $\alpha_{\pm 2},\alpha_{\pm 1},\alpha_0$ given by Eqs. 
(\ref{comp-3})-(\ref{comp-4})  
the coefficient $\Theta$ of Eq. (\ref{theta}) is identically equal to 0. The condition $\Theta=0$ 
for the ground state
makes the 
GP equations (\ref{gp_s1})-(\ref{gp_s3})  simpler and independent of $c_2$. Consequently,
the DM equation (\ref{DM-1})  becomes an  exact equation for the ground state wave function provided Eq. (\ref{comp-2}) holds, e.g. the component wave functions are multiples of each other.
Our numerical calculations show that the condition (\ref{comp-2}) holds 
for all magnetization to a very high degree of accuracy. 

The coefficients $\alpha_{\pm 2},\alpha_{\pm 1},\alpha_0$ can also be obtained from a minimization
of  energy 
along with condition (\ref{comp-2}),
for a ferromagnetic ground state with $c_1<0$ and $c_2>20c_1$.  An explicit account of the derivation of the coefficients 
$\alpha_{\pm 2},\alpha_{\pm 1}, \alpha_0,$ from an  energy minimization for a ferromagnetic spin-2 
BEC is given in  Appendix B. 

%\int_{-l_{\pm 1}}^ {l_{\pm 1}}\rho_{\pm 1}(x) d{\bf x} = \frac{ 1\pm m}{2}.
For an  anti-ferromagnetic  BEC ($c_2<0, c_2<20c_1$) for
any non-zero magnetization $m$ numerical studies show that 
$\phi_{\pm 1}(\mathbf x)=\phi_0(\mathbf x)=0$ for the ground state.
This can also be  obtained from energetic consideration  as  shown in Appendix B. 
The   magnetization  and normalization conditions (\ref{mag}) and (\ref{nor})
then yield
\begin{align}\label{dist-3}
\int \rho_{\pm 2} d{\bf x} = \frac{2\pm m}{4},
\end{align} 
Energy consideration establishes that
a cyclic BEC ($c_1>0,c_2>0$)  
has two 
degenerate ground states for all non-zero magnetization $m$ with $(i) \phi_{+1}(\mathbf x) = 
\phi_{0}(\mathbf x)=\phi_{-2}(\mathbf x)=0$, or with 
$(ii) \phi_{\pm 1}(\mathbf x)=0,$ viz. Appendix B. Consequently,  the magnetization and normalization conditions (\ref{mag}) and (\ref{nor})
lead for these two states 
\begin{align}\label{dist-5}
&(i) \int  d{\bf x} \rho_{+2}= \frac{1+m}{3}, \quad  \int  d{\bf x}
  \rho_{-1}= \frac{2-m}{3},\\
&(ii) \int  d{\bf x} \rho_{\pm 2}= \left(\frac{2\pm m}{4}\right)^2, \quad \int  d{\bf x} \rho_{0}= \frac{4-m^2}{8}.
\label{dist-6}
\end{align} 
%{ here $l_{\pm j}$ with $j=2,1,0,-1,-2$ are the spatial extensions of $m_f = \pm j$ components.}
%\int_{-l_{\pm 1}}^ {l_{\pm 1}}\rho_{\pm 1}(x) d{\bf x} = \frac{ 1\pm m}{2}.
 
For both  anti-ferromagnetic and cyclic BECs we will derive the analytic models directly from the TFA to the GP equations 
 (\ref{gp_s1})-(\ref{gp_s3}) and not from the DM, whereas for a ferromagnetic BEC we will 
rely on the TFA to the DM equation (\ref{DM-1}) with $C=(c_0+4c_1).$

%%%%%%%%%%%%%%%%%%%%%%%%%%%%%%%%%%%%%%%%%%%%%%%%%%%%%%%%%%%%%%%%%%%%%%%%%%%%%%%%%%%%

\section{Analytic model for spin-1 BEC}

\label{Sec-III}

\subsection{Ferromagnetic BEC}
\label{IIA}

\begin{figure}[!t]
\begin{center}
\includegraphics[trim = 0mm 0mm 0cm 0mm, clip,width=1.1\linewidth,clip]{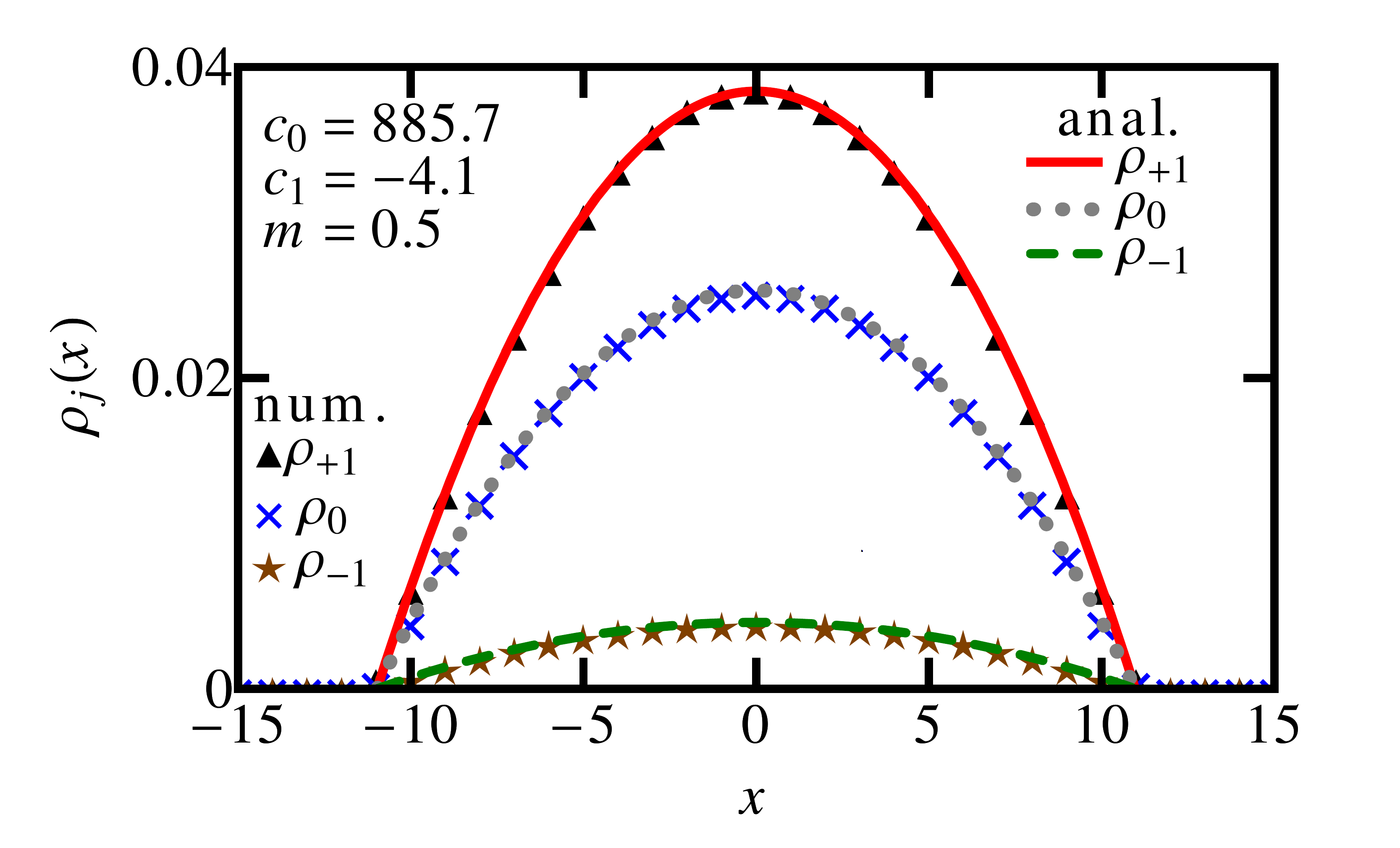}
\caption{(Color online) Analytic (anal.) and numerical (num.) densities of a spin-1 quasi-1D ferromagnetic
$^{87}$Rb BEC. The number of atoms, scattering lengths and oscillator lengths
are, respectively, $N = 10,000$, $a_0 = 101.8a_B$, $a_2 = 100.4a_B$ \cite{kempen}, 
$l_0 = 2.41\mu$m, $l_{yz} = 0.54\mu$m, here $a_B$ is Bohr radius.}
\label{fig-3} \end{center}
\end{figure}

\begin{figure}[!t]
\begin{center}
\includegraphics[trim = 0mm 0mm 0cm 0mm, clip,width=1.1\linewidth,clip]{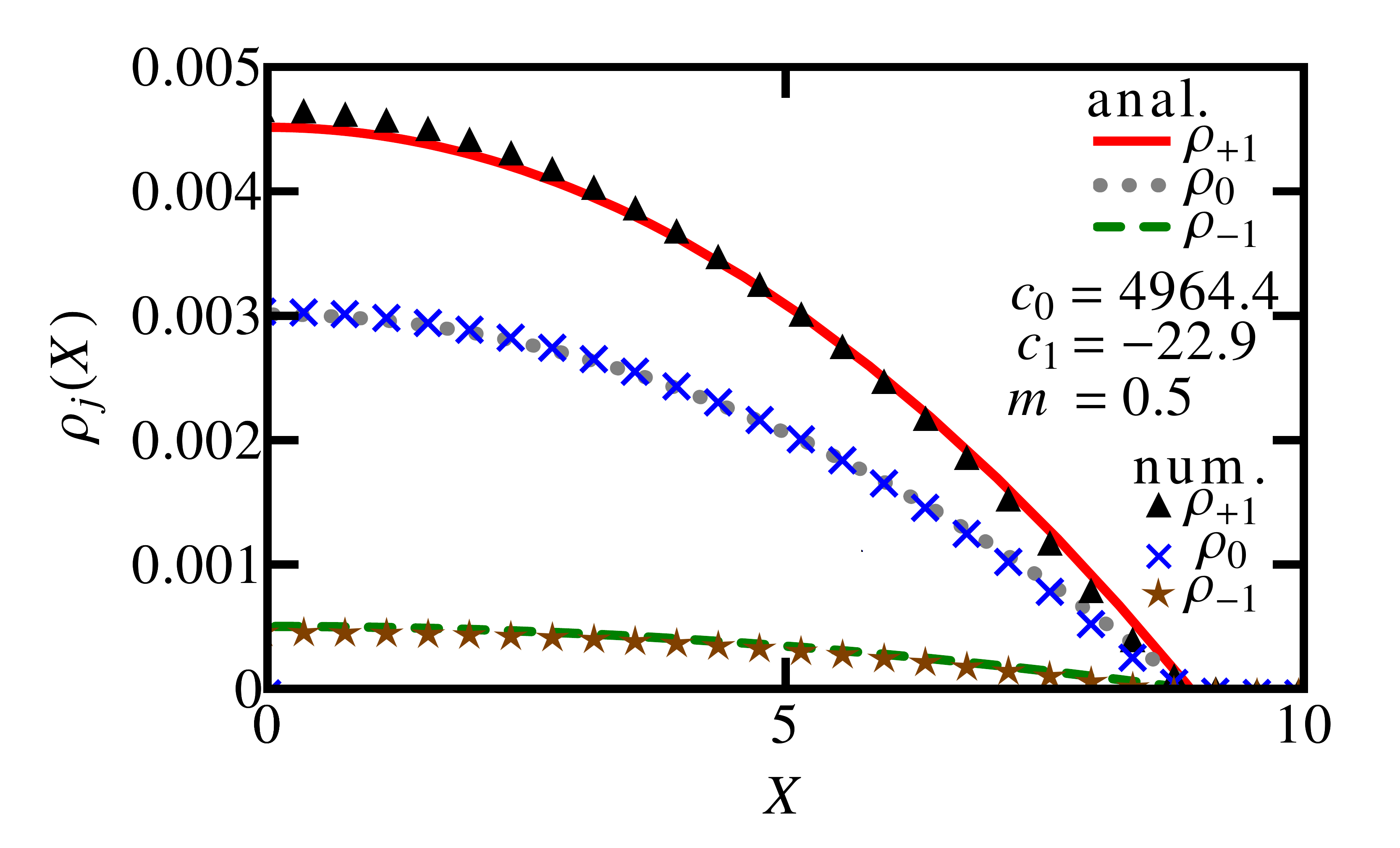}
\caption{(Color online) Analytic (anal.) and numerical (num.) densities of a spin-1 quasi-2D ferromagnetic
$^{87}$Rb BEC. The number of atoms, scattering lengths and oscillator lengths
are, respectively, $N = 100,000$, $a_0 = 101.8a_B$, $a_2 = 100.4a_B$ \cite{kempen}, {
$l_0 = 2.41\mu$m,} $l_{z} = 0.54\mu$m, here $a_B$ is Bohr radius.}
\label{Fig-3} \end{center}
\end{figure}

We derive the analytic model for the ground-state density of a spin-1 BEC 
using the TFA to  the   DM equation (\ref{DM-1}) with component densities given by Eq. (\ref{dist-1}). In the  TFA the kinetic energy term in Eq.  (\ref{DM-1}) 
  is neglected, which is reasonable for a moderate to large positive nonlinear terms, and the BEC density is calculated by  equating the ``Hamiltonian'' to the chemical potential by
\begin{equation}\label{dmtfa}
   \mu= [{\tiny  X}^2/2+C\rho_{\mathrm{DM}}], \quad C\equiv {\cal C}_I=c_0+c_1,
\end{equation}
thus leading to the 
  TFA density 
\begin{align}\label{tfaferro}
\rho_{\mathrm{DM}}(X)={{(l_{\cal D}^2-X^2)}}/{(2{\cal C}_I)}, \quad {X{\le} l_{\cal D}=}  \sqrt{2\mu}.
\end{align}
% where ${\cal D}=1,2,3$
%is the dimensionality of space.
Imposing the condition of normalization    ${\int \rho_{\mathrm{DM}}(X)d{\bf x}}=1$, we obtain, in 1D, 2D, and 3D, respectively 
$l_{\cal D}=(3{\cal C}_I/2)^{1/3}, (4{\cal C}_I/\pi)^{1/4}$ and $(15{\cal C}_I/4\pi)^{1/5}$, 
%{\begin{align}\label{tfaferro}
% \rho_{\mathrm{DM}}(x)=&\frac{ (3C/2)^{2/3}-x^2}{2C}, 
%\quad x{\le}  \sqrt{2\mu}\equiv  \left(\frac{3C}{2}\right)^{1/3},\\
% \rho_{\mathrm{DM}}(x)=&\frac{ (4C/\pi)^{1/2}-x^2}{2C}, 
%\quad x{\le}  \sqrt{2\mu}\equiv  \left(\frac{4C}{\pi}\right)^{1/4},\\
%\rho_{\mathrm{DM}}(x)=&\frac{ (15C/4\pi)^{2/5}-x^2}{2C}, 
%\quad x{\le}  \sqrt{2\mu}\equiv  \left(\frac{15C}{4\pi}\right)^{1/5},
%\end{align}}
provided ${\cal C}_I>0$.
The component densities are calculated using  Eqs.  (\ref{sma}) and (\ref{dist-1}).
%where $\mu = l^2/2$ with $l =  [3(c_0+c_2)/2]^{1/3}$ 
%denoting the spatial extent of the BEC. 
The analytic  densities  for a quasi-1D spin-1 ferromagnetic $^{87}$Rb BEC  are shown in Fig. \ref{fig-3} along with the 
numerical solution of the full coupled GP equations (\ref{gps-1})-(\ref{gps-2}).
The same for a quasi-2D spin-1 ferromagnetic $^{87}$Rb BEC is shown in 
Fig. \ref{Fig-3}.
All numerical calculations are performed using the split-step Crank-Nicolson 
scheme \cite{cn} with space and time steps 0.025  and 0.00005, respectively.

{ 
In the DM model ${\cal C}_I\sim a_2$ plays the same role as the scattering length $a$ 
in a scalar BEC in Eq. (\ref{tfasize}). Hence in this case the condition of validity 
of the TFA will be $Na_2/l_0 >> 1$. 
}
\subsection{Anti-ferromagnetic BEC}

\label{IIB}

\begin{figure}[!t]
\begin{center}
\includegraphics[trim = 0mm 0mm 0cm 0mm, clip,width=1.1\linewidth,clip]{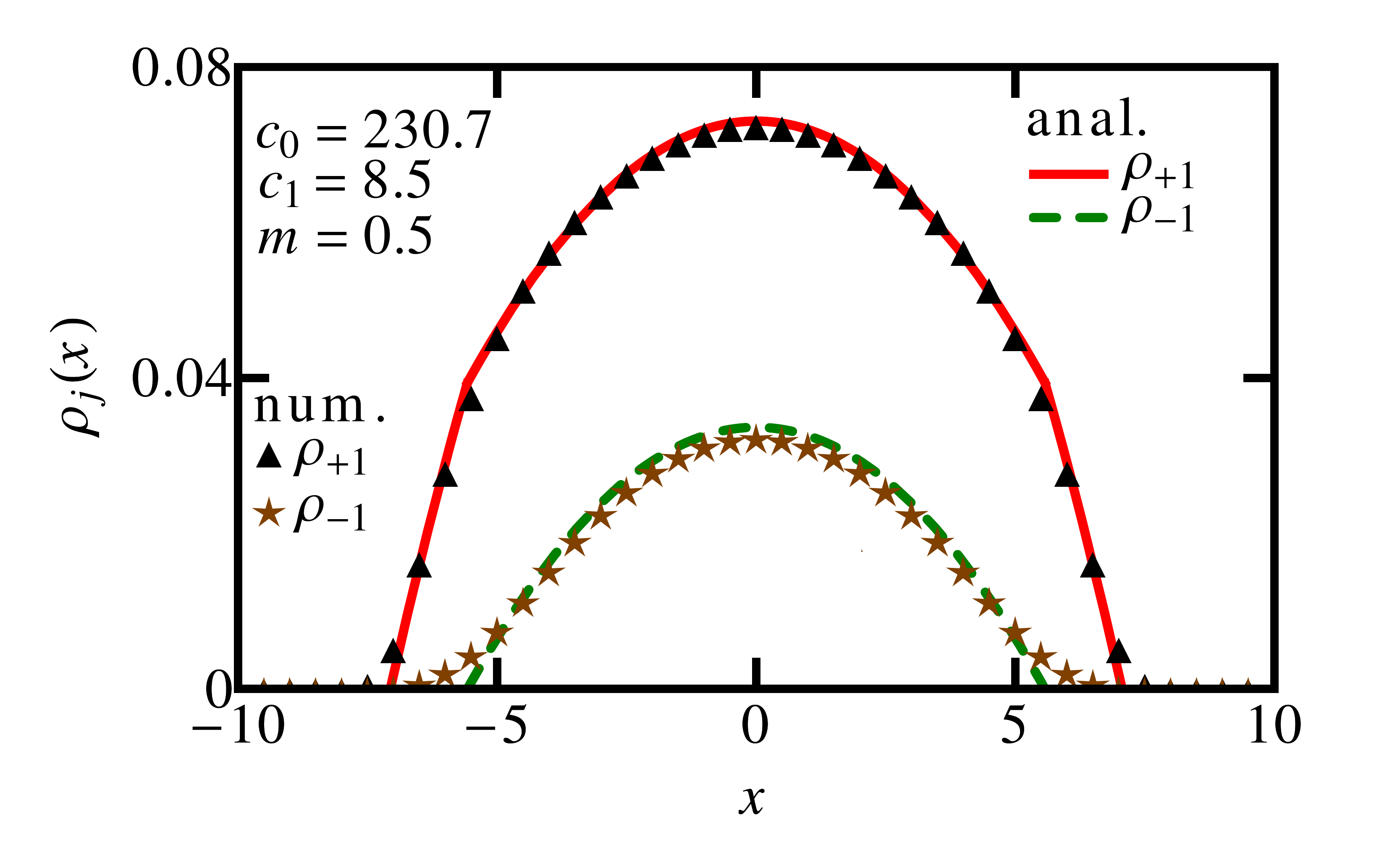}
\caption{(Color online) Analytic (anal.) and numerical (num.)
densities of a spin-1 quasi-1D anti-ferromagnetic
$^{23}$Na BEC.
The number of atoms, scattering lengths and oscillator lengths
are, respectively, $N = 10,000$; $a_0 = 47.36a_B$, $a_2 = 52.98a_B$ \cite{ueda}; 
$l_0 = 4.69\mu$m, $l_{yz} = 1.05\mu$m.
}
\label{fig-2} \end{center}
\end{figure}

\begin{figure}[!t]
\begin{center}
\includegraphics[trim = 0mm 0mm 0cm 0mm, clip,width=1.1\linewidth,clip]{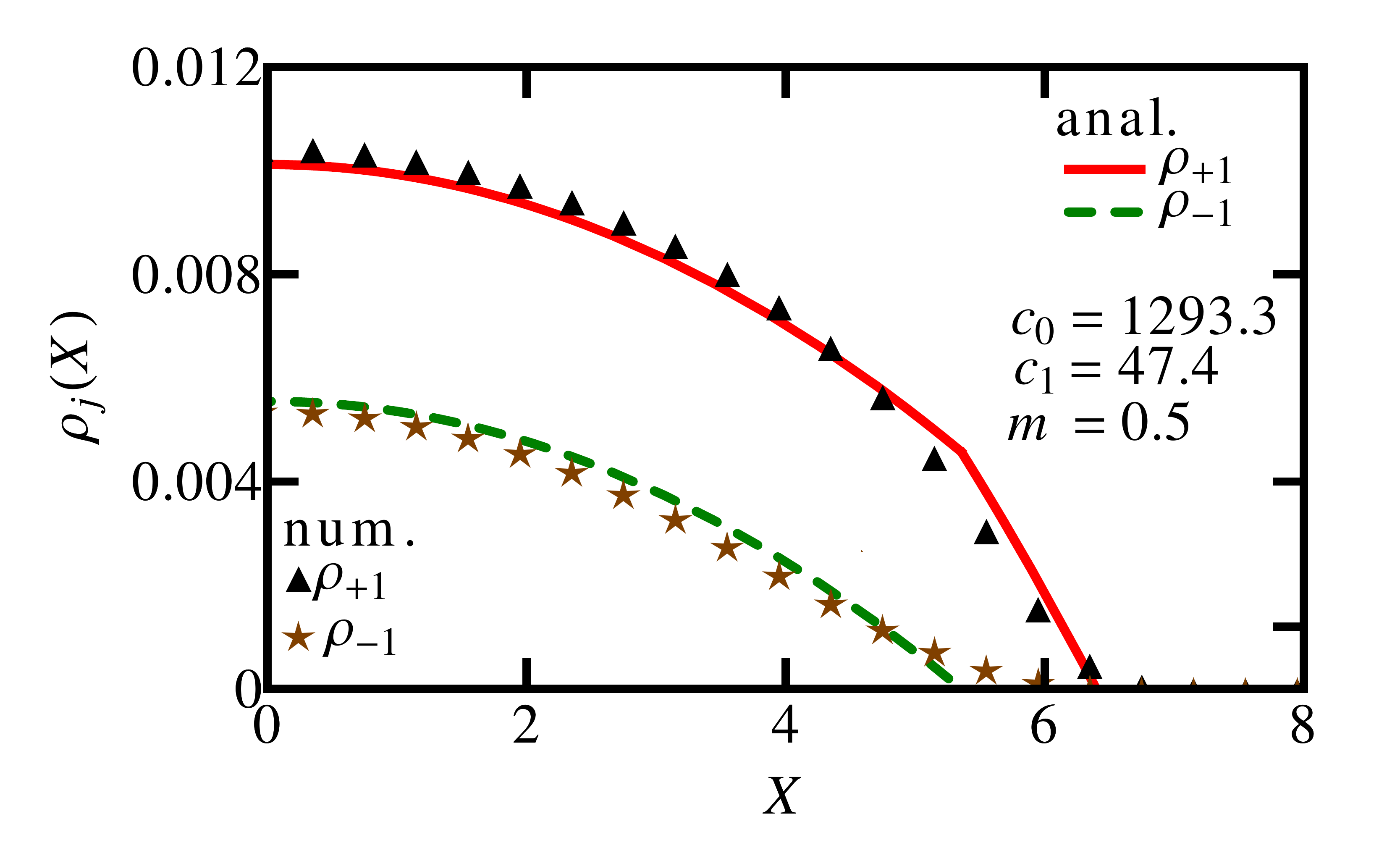}
\caption{(Color online) Analytic (anal.) and numerical (num.)
densities of a spin-1 quasi-2D anti-ferromagnetic
$^{23}$Na BEC.
The number of atoms, scattering lengths and oscillator lengths
are, respectively, $N = 100,000$; $a_0 = 47.36a_B$, $a_2 = 52.98a_B$ \cite{ueda}; {
$l_0 = 4.69\mu$m,} $l_{z} = 1.05\mu$m.
}
\label{Fig-2} \end{center}
\end{figure}

In this case, the analytic model is derived by applying TFA directly to the GP equation 
(\ref{gps-1}) 
 with $\phi_0({\mathbf x})=0$. 
For a non-zero  magnetization ($0<m<1$), the ${m_f}=+1$ component accommodates more atoms and  
its spatial extension $ l_{{\cal D}(+1)}$ is larger than  the same of the  the ${m_f}=-1$ component with 
spatial extension $ l_{{\cal D}(-1)}$. Hence for $l_{{\cal D}(+1)}>x>l_{{\cal D}(-1)}$, $\phi_{-1}({\bf x})=0$  and the coupled GP equation 
(\ref{gps-1})  for $\phi_{\pm 1}({\bf x})$
reduces to a single equation for $\phi_{+ 1}({\bf x})$.
In the TFA,  the kinetic energy
terms in the GP equation (\ref{gps-1}) are neglected and the densities are calculated by equating 
the Hamiltonian to the respective chemical potentials: 
\begin{align}\label{TF_eqs_polar}
 \mu_{\pm 1}  &=
 \left[ X^2/2+{c}_0{\rho}\right]
+c_1({\rho}_{\pm 1} -{\rho}_{\mp 1}),\quad X{\le }l_{{\cal D}(-1)},\\
 \mu_{+ 1} &=\left[ X^2/2+{\cal C}_I{\rho}_{+1}\right]
,\quad  l_{{\cal D}(+1)}{\ge }X {\ge }l_{{\cal D}(-1)},
\label{TF_eqs_polar2}
\end{align}
  subject to   normalization (\ref{norm-1}). 
In the domain $l_{{\cal D}(+1)}{ \ge }X{ \ge }l_{{\cal D}(-1)} $,   Eq. (\ref{TF_eqs_polar2}) 
has the solution
\begin{align}\label{x3}
\rho_{+1}(X)=& \frac{l_{{\cal D}(+1)}^2-X^2}{2{\cal C}_I}, \quad l_{{\cal D}(+1)}{ \ge }X{ \ge }l_{{\cal D}(-1)},\\
\mu_{+1}=& l_{{\cal D}(+1)}^2/2. \label{x2}
\end{align}
In the overlap region  $X{ \le }l_{{\cal D}(-1)}$, coupled equations  (\ref{TF_eqs_polar}) have the solution
\begin{align}
\rho_{\pm 1}(X) &= \frac{c_0 ( \mu_{\pm 1} - \mu_{\mp 1})+c_1 
(\mu_{+1}+\mu_{-1}-X^2)}{4 c_0 c_1}\label{rho_p1}. 
\end{align}
The condition $\rho_{-1}(l_{{\cal D}(-1)})=0$ leads to 
\begin{eqnarray}
\mu_{-1} &=& \frac{(c_0-c_1) l_{{\cal D}(+1)}^2+2 c_1 l_{{\cal D}(-1)}^2}{2 {\cal C}_I}.\label{mu_m1}
\end{eqnarray} 
Substituting Eqs. (\ref{x2}) and (\ref{mu_m1}) in Eq. (\ref{rho_p1}),   we obtain  
\begin{align}\label{dens-1}
\rho_{+1}(X)                =& { \frac{2 c_0 l_{{\cal D}(+1)}^2+(c_1 -c_0)l_{{\cal D}(-1)}^2-{\cal C}_I X^2}{4 c_0 {\cal C}_I}, \quad  X{\le }l_{{\cal D}(-1)}} \\
\rho_{-1}(X)&=
            {\frac{(l_{{\cal D}(-1)}^2-X^2)}{4 c_0}, \quad X{\le }l_{{\cal D}(-1)}}.\label{dens-2}
\end{align}

The normalization condition (\ref{norm-1}) for $\rho_{\pm1}(X)$
  leads 
 to 
\begin{eqnarray}
%l_{{\cal D}(+1)} &=&  [3(c_0+c_1 m)/2]^{1/3}\label{l_1},\\
l_{{\cal D}(-1)}&=& l_{\cal D} [c_0(1- m)/{\cal C}_I]^{1/(2+{\cal D})},\label{l_m1d}\\
%l_{{ 2}(-1)}&=&  [4c_0(1- m)/\pi]^{1/4},\label{l_m2d}\\
%l_{{ 3}(-1)}&=&   [15 c_0(1- m)/(4\pi)]^{1/5} ,\label{l_m3d}
%\end{eqnarray}
%and the same for $\rho_{+1}(x)$ leads to
%\begin{eqnarray}
l_{{{\cal D}}(+1)} &=& l_{\cal D} [(c_0+c_1 m)/{\cal C}_I]^{1/(2+{\cal D})}\label{l_p1d}.
%l_{{\cal D}(+1)}&=&  [3c_0(1- m)/2]^{1/3},\label{l_m1}\\
%l_{{ 2}(+1)}&=&  [4(c_0+c_1 m)/\pi]^{1/4},\label{l_p2d}\\
%l_{{ 3}(+1)}&=& [15(c_0+c_1 m)/(4\pi)]^{1/5} \label{l_p3d},
\end{eqnarray}
%Integrating Eqs. (\ref{TF_eqs_polar}) and (\ref{TF_eqs_polar2})
%from $-l_{{\cal D}(+1)}$ to $l_{{\cal D}(+1)}$ and from $-l_{{\cal D}(-1)}$ to $l_{{\cal D}(-1)}$ for $\mu_{+1}$ and $\mu_{-1}$, respectively,
%{and normalizing the densities
%according to Eq. (\ref{norm1})}, 
%we get 
%\begin{align}
%2\mu_{+ 1} l_{+ 1} =& \frac{l_{{\cal D}(+1)}^3}{3} + c_0 + c_1m,\label{mu}\\
%{2\mu_{-1} l_{{\cal D}(-1)}} =& {\frac{l_{{\cal D}(-1)}^3}{3}  +(c_0 +c_1) \frac{1-m}{2} +(c_0-c_1)}%\nonumber\\
%  & {\times \frac{3 c_0 l_{{\cal D}(+1)}^2 l_{{\cal D}(-1)}-2 c_0 l_{{\cal D}(-1)}^3+c_1 l_{{\cal D}(-1)}^3}{3 c_0^2+3 c_0 c_1}}. \label{qu}
%\end{align}
%The positive roots of Eqs. (\ref{mu}) and (\ref{qu}) are
%\begin{eqnarray}
%l_{{\cal D}(+1)} &=&  [3(c_0+c_1 m)/2]^{1/3}\label{l_1},\\
%l_{{\cal D}(-1)}&=&  [3c_0(1- m)/2]^{1/3}.\label{l_m1}
%\end{eqnarray}
  The   densities   (\ref{x3}), (\ref{dens-1}) and (\ref{dens-2}) with 
extensions given by Eqs. (\ref{l_m1d})-(\ref{l_p1d}) constitute the analytic model in this case.

These analytic densities for a spin-1 quasi-1D
 anti-ferromagnetic
$^{23}$Na BEC are shown in Fig. \ref{fig-2} along with the numerical solution of the full coupled
GP equations (\ref{gps-1})-(\ref{gps-2}). 
The same for a spin-1 quasi-2D  anti-ferromagnetic
$^{23}$Na BEC are shown in Fig. \ref{Fig-2}. 
Comparing Eq. (\ref{tfasize}) with Eqs. (\ref{l_m1d})-(\ref{l_p1d}), the conditions for the validity of TFA in this case are
\begin{align}
\frac{N(a_0+2a_2)(1-m)}{3l_0}&>>1,\label{tfapolarp1}\\
\frac{N[a_0+2a_2+(a_2-a_0)m]}{3l_0}&>>1\label{tfapolarm1},
\end{align}
for $m_f=-1$ and $m_f=+1$ component, respectively. The terms on
the left side of Eqs. (\ref{tfapolarp1}) and (\ref{tfapolarm1})
are the measure of the repulsive interactions in $m_f=-1$
and $m_f=+1$ components, respectively.

For magnetization $m=0$ there is another degenerate ground state 
where all atoms are in the $m_f=0$ component \cite{Kawaguchi}. In that case the 
spin-1 GP equation reduces to the DM equation (\ref{dmtfa}) with ${\cal C}_I
=c_0$ and  
  $\rho_0(X)=\rho_{DM}(X)$ of Eq. (\ref{tfaferro}). From Eqs. (\ref{tfapolarp1})-(\ref{tfapolarm1}), 
the simple criterion for the validity of TFA in this case is $N(a_0+2a_2)/(3l_0)>>1$, which is consistent with
the fact that for $m=0$, $c_1$ term does not contribute to the energy of the system.

The TF analysis shows that the spatial extents of the components are different
for an anti-ferromagnetic BEC for any non-zero magnetization, viz. 
 Eqs. (\ref{l_m1d})-(\ref{l_p1d}), which is also manifested by different 
chemical potentials, viz. Eqs. (\ref{x2}) and (\ref{mu_m1}), whereas these are the same for 
a ferromagnetic BEC. Thus, in the domain $l_{{\cal D}(-1)}{\le }X<l_{{\cal D}(+1)}$ only component 
${m_f} = +1$ survives for the anti-ferromagnetic BEC effectively separating 
this component from mixed phase in the $X<l_{{\cal D}(-1)}$ domain.  The different spatial
extents of the components for an anti-ferromagnetic BEC also imply that SMA is not
valid in general except for $m=0$, when $l_{{\cal D}(\pm 1)}$ of 
{Eqs. (\ref{l_m1d})-(\ref{l_p1d})
become equal. {The ground states shown in Figs. \ref{fig-2} and 
\ref{Fig-2} preserve the symmetry of the trapping potential. These symmetric
profiles minimize $c_1$-dependent interaction energy \cite{Yi}
\begin{align}
E_{\rm A} =& \frac{c_1}{2}\int \big[\left( \rho_{+1}-\rho_{-1}\right)^2+2\rho_0(\rho_{+1}+\rho_{-1}) \nonumber \\
          &-4\sqrt{\rho_{+1}\rho_{-1}}\rho_0\big]d{\bf x}.
\end{align}
The asymmetric states, where the two phase-separated components lie side by
side \cite{sadhan}, do not minimize $E_{\rm A}$ in addition to {having more potential-energy contribution}. Hence,
they do not emerge as the ground states in trapped spinor condensates. The
asymmetric states can emerge as the ground state in the presence of Zeeman energy 
\cite{Kawaguchi,Matuszewski} or spin-orbit coupling \cite{sadhan,GA-2} which we do not include in the Hamiltonian.}

%%%%%%%%%%%%%%%%%%%%%%%%%%%%%%%%%%%%%%%%%%%%%%%%%%%%%%%%%%%%%%%%%%%%%%%%%%%%%%%%%%%%%%%%%%%%%%%%%%%%%%%%%%%%%%%%%%%%%%%%%%%%%%%%%%%%%%%%%%%%%

\section{Analytic model  for spin-2 BEC} 

\label{Sec-IV}

\subsection{Ferromagnetic BEC}

\begin{figure}[!t]
\begin{center}
\includegraphics[trim = 0mm 0mm 0cm 0cm, clip,width=1.1\linewidth,clip]{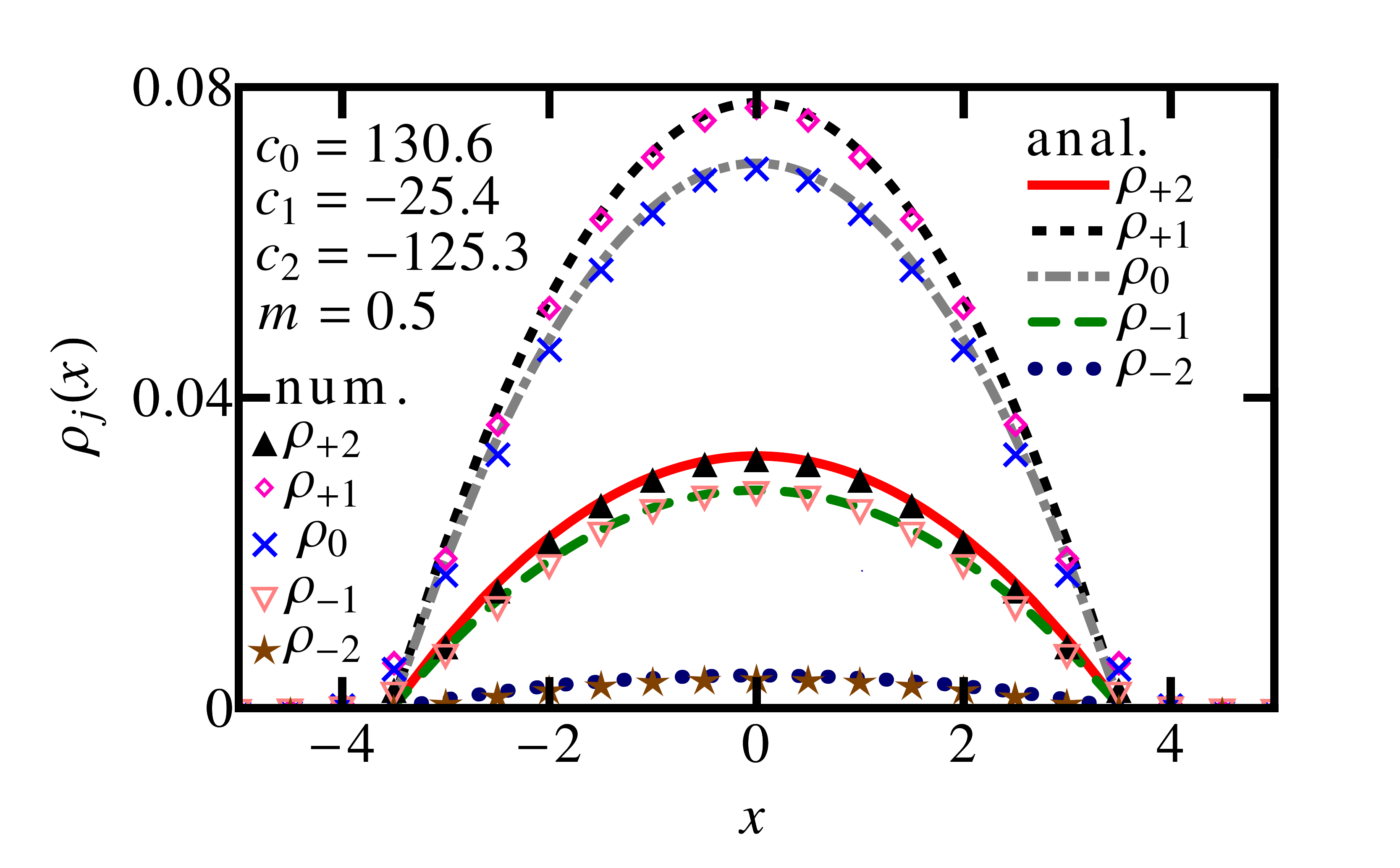}
\includegraphics[trim = 0mm 0mm 0cm 0cm, clip,width=1.1\linewidth,clip]{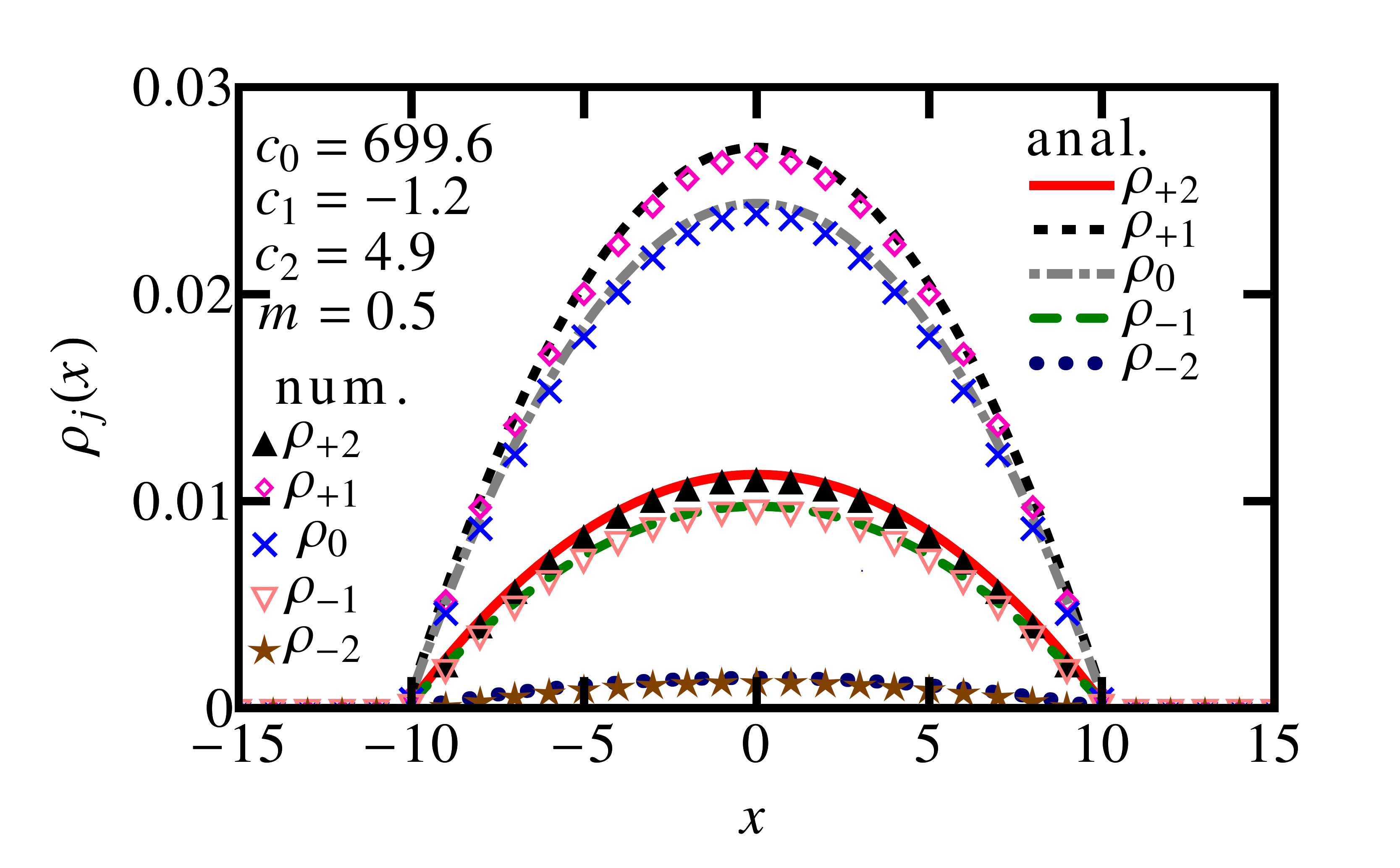}
\caption{(Color online) (a) and (b) show 
analytic (anal.) and numerical (num.) densities of quasi-1D spin-2 
ferromagnetic $^{23}$Na and $^{83}$Rb BECs, respectively.
For $^{23}$Na, the number of atoms, scattering lengths and oscillator lengths
are, respectively, $N = 10,000$, $a_0 = 34.9a_B$, $a_2 = 45.8a_B$, $a_4 = 6.45a_B,$
\cite{Ciobanu} $l_0 = 4.69\mu$m, $l_{yz} = 1.05\mu$m,
here $a_B$ is Bohr radius. The experimental value of $a_4  (=64.5a_B)$ has been modified to access 
the ferromagnetic phase of $^{23}$Na  (using a Feshbach resonance) from the natural anti-ferromagnetic phase. 
For $^{83}$Rb, the corresponding parameters are $N = 10,000$, $a_0 = 83.0a_B$, $a_2 = 82.0a_B$, $a_4 = 81.0a_B,$
\cite{Ciobanu} $l_0 = 2.47\mu$m, $l_{yz} = 0.55\mu$m.}
\label{fig-6} \end{center}
\end{figure}

In this case, the component densities are given by the DM equation  (\ref{comp-2})
along with distributions  
(\ref{comp-3})-(\ref{comp-4}) and the analytic model is derived 
from the TFA to the  DM equation (\ref{DM-1}).
Following the procedure  of Sec. \ref{IIA} for a spin-1 ferromagnetic BEC, 
the TFA densities are again given by Eq. (\ref{tfaferro}), but now with $C\equiv{\cal C}_{II}=c_0+4c_1$.
The component densities are then obtained using Eqs. (\ref{comp-3})-(\ref{comp-4}).
%The analytic  and numerical densities for $c_0=130.6,c_1=-25.4, c_2=-125.3$ are shown in Fig. \ref{fig-6}.
These analytic densities for a quasi-1D spin-2 ferromagnetic $^{23}$Na {and $^{83}$Rb BECs are shown in Figs.  \ref{fig-6}(a) and
(b), respectively,}
together with the numerical densities from the full coupled GP equations (\ref{gp_s1})-(\ref{gp_s3}).

In the DM model ${\cal C}_{II}\sim a_4$ plays the same role as the scattering length $a$ 
in a scalar BEC in Eq. (\ref{tfasize}). Hence in this case the condition of validity 
of the TFA will be $Na_4/l_0 >> 1$.

\begin{figure}[!t]
\begin{center}
\includegraphics[trim = 0mm 0mm 0cm 0mm, clip,width=1.1\linewidth,clip]{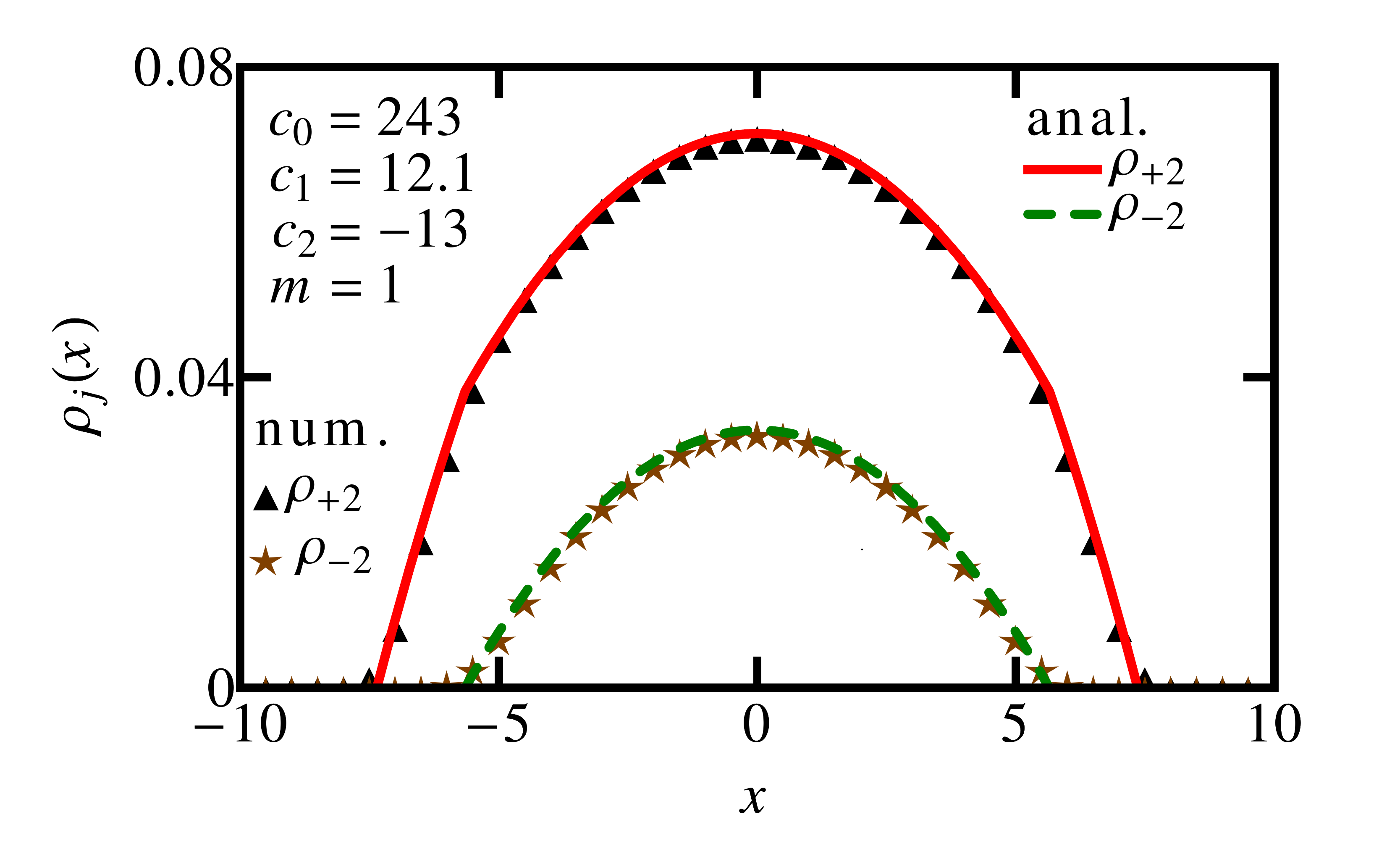}
\caption{(Color online)  Analytic (anal.) and numerical (num.) densities of a spin-2 quasi-1D anti-ferromagnetic  $^{23}$Na BEC. The number of atoms, scattering lengths and oscillator lengths
are, respectively, $N = 10,000$; $a_0 = 34.9a_B$, $a_2 = 45.8a_B$, $a_4 = 64.5a_B$ \cite{Ciobanu} ; $l_0 = 4.69\mu$m, $l_{yz} = 1.05\mu$m.}
\label{fig-7} \end{center}
\end{figure}

\subsection{Anti-ferromagnetic BEC}

The analytic model here is obtained from the TFA to the GP equation 
(\ref{gp_s1}) involving only $\phi_{\pm 2}({\mathbf x})$
subject to $\phi_0({\mathbf x})=\phi_{\pm 1}({\mathbf x})=0$ with  normalization condition (\ref{dist-3}).
 After neglecting the kinetic energy terms in the GP equation   (\ref{gp_s1})  the 
corresponding TFA densities are   given by  
\begin{align}&
 \mu_{\pm 2} =  X^2/2+c_0 \rho \pm 4 c_1 (\rho_{+2} - \rho_{-2})\label{gps_polar_1}
   +\frac{2c_2 \rho_{\mp 2}}{5}, 
\end{align}
For a non-zero magnetization $0<{ m}\le 2$, the ${m_f}=+2$ component extends to a larger domain 
($X<l_{{\cal D}(+2)}$) than the ${m_f}= -2$ component
with a smaller extension ($X<l_{{\cal D}(-2)}, l_{{\cal D}(+2)}>l_{{\cal D}(-2)}$).
Following the procedure presented in Sec. \ref{IIB} for a spin-1 anti-ferromagnetic BEC, one can calculate  $l_{{\cal D}(+2)}$ 
in 1D, 2D and 3D, respectively, as
\begin{align}
l_{{1}(+2)} &=& \frac{[3{\cal A}]^{1/3}}{20^{1/3}},\label{lp2_spin2_polar}\quad
l_{{2}(+2)} =\frac{[2{\cal A}]^{1/4}}{(5 \pi )^{1/4}},\quad
l_{{3}(+2)} = \frac{[3{\cal A}]^{1/5}}{(8\pi) ^{1/5}},
\end{align}
where ${\cal A}=\left(10 c_0+ 2 c_2+20 c_1 m-c_2 m\right).$
Similarly, $l_{{\cal D}(-2)}$ in 1D, 2D and 3D, respectively, are
\begin{align}
l_{{ 1}(-2)} = \frac{ [3{\cal B}]^{1/3} }{20^{1/3}}\label{lm2_spin2_polar},\quad
l_{{ 2}(-2)} = \frac{[2 {\cal B}]^{1/4}}{(5 \pi )^{1/4}},\quad
l_{{ 3}(-2)} = \frac{[3{\cal B}]^{1/5} }{(8\pi)^{1/5}},
\end{align}
where ${\cal B}= (5 c_0+c_2)(2-m)$.
The analytic TFA densities of the spin components $m_f=\pm 2$ are given by 
\begin{align}\label{rhop2_spin2_polar}
\rho_{+2}(X) &= \frac{20 c_1 \left(l_{{\cal D}(-2)}^2-X^2\right)+2 c_2 \left(l_{{\cal D}(+2)}^2-l_{{\cal D}(-2)}^2\right)  }
            {4 {\cal C}_{II} (5 c_0+c_2)}
             \nonumber \\
&+ \frac{ 5 c_0 \left(2 l_{{\cal D}(+2)}^2-l_{{\cal D}(-2)}^2-X^2\right) }
            {4 {\cal C}_{II} (5 c_0+c_2)},
             \quad X{\le }l_{{\cal D}(-2)},\\
%\end{align}
%\begin{align}
    \rho_{+2}(X)      &= \frac{l_{{\cal D}(+2)}^2-X^2}{2{\cal C}_{II}}, \quad l_{{\cal D}(+2)}{\ge }X{\ge }l_{{\cal D}(-2)},\label{rhop2spin2_polar}\\
\rho_{-2}(X) &= \frac{5 \left(l_{{\cal D}(-2)}^2-X^2\right)}{4 (5 c_0+c_2)}, \quad X{\le }l_{{\cal D}(-2)}\label{rhom2spin2_polar},
\end{align}
with the extensions $l_{\pm 2}$ given by Eqs. (\ref{lp2_spin2_polar}) and 
(\ref{lm2_spin2_polar}).
The analytic  and numerical  densities  for a quasi-1D spin-2 anti-ferromagnetic   $^{23}$Na BEC
 are compared in Fig. \ref{fig-7}. {Again, in this case, phase-separated asymmetric profiles
do not emerge {as ground states}
due to more energy contribution from $c_2$-dependent energy term in
addition to more potential energy as compared to symmetric profiles.

Comparing Eq. (\ref{tfasize}) with Eqs. (\ref{lp2_spin2_polar})-(\ref{lm2_spin2_polar}), 
the conditions for the validity of TFA in this case are
\begin{align}
%\frac{N[4a_2+3a_4+2(a_4-a_2)m+(7a_0-10a_2+3a_4)(2-m)/10]}{7l_0}&>>1,\label{tfaspin2afp2}\\
%\frac{N[(7a_0+10a_2+18a_4)(2-m)/10]}{7l_0}&>>1,\label{tfaspin2afm2}\\
\frac{N[c_{0a}+2c_{1a}m+c_{2a}(2-m)/10]}{7l_0}&>>1,\label{tfaspin2afp2}\\
\frac{N[(c_{0a}+c_{2a}/5)(1-m/2)]}{7l_0}&>>1,\label{tfaspin2afm2}
\end{align}
for $m_f=+2$ and $m_f=-2$ components, respectively, where $c_{0a} = 4a_2+3a_4$, $c_{1a} = a_4-a_2$,
and $c_{2a} = 7a_0-10a_2+3a_4$.  
} 

For $m=0$ there is another  degenerate ground  state with the all the atoms in the 
$m_f=0$ component \cite{ueda}. In this case  the GP equation reduces to the DM equation (\ref{dmtfa})
with ${\cal C}_{II}=(c_0+c_2/5)$ and $\rho_{0}(X)=\rho_{DM}(X)$ of 
(\ref{tfaferro}). A superposition of this solution  and the solution corresponding to 
Eq. (\ref{gps_polar_1}) with $m=0$ will also be a degenerate solution.
The simpler criterion for the validity of TFA in this case is 
$N[(7a_0+10a_2+18a_4)/5]/(7l_0)>>1$, which is
consistent with Eqs. (\ref{tfaspin2afp2}) and (\ref{tfaspin2afm2}) with $m=0$.

\subsection{Cyclic BEC}

\begin{figure}[!b]
\begin{center}
\includegraphics[trim = 0mm 0mm 0cm 0mm, clip,width=1.1\linewidth,clip]{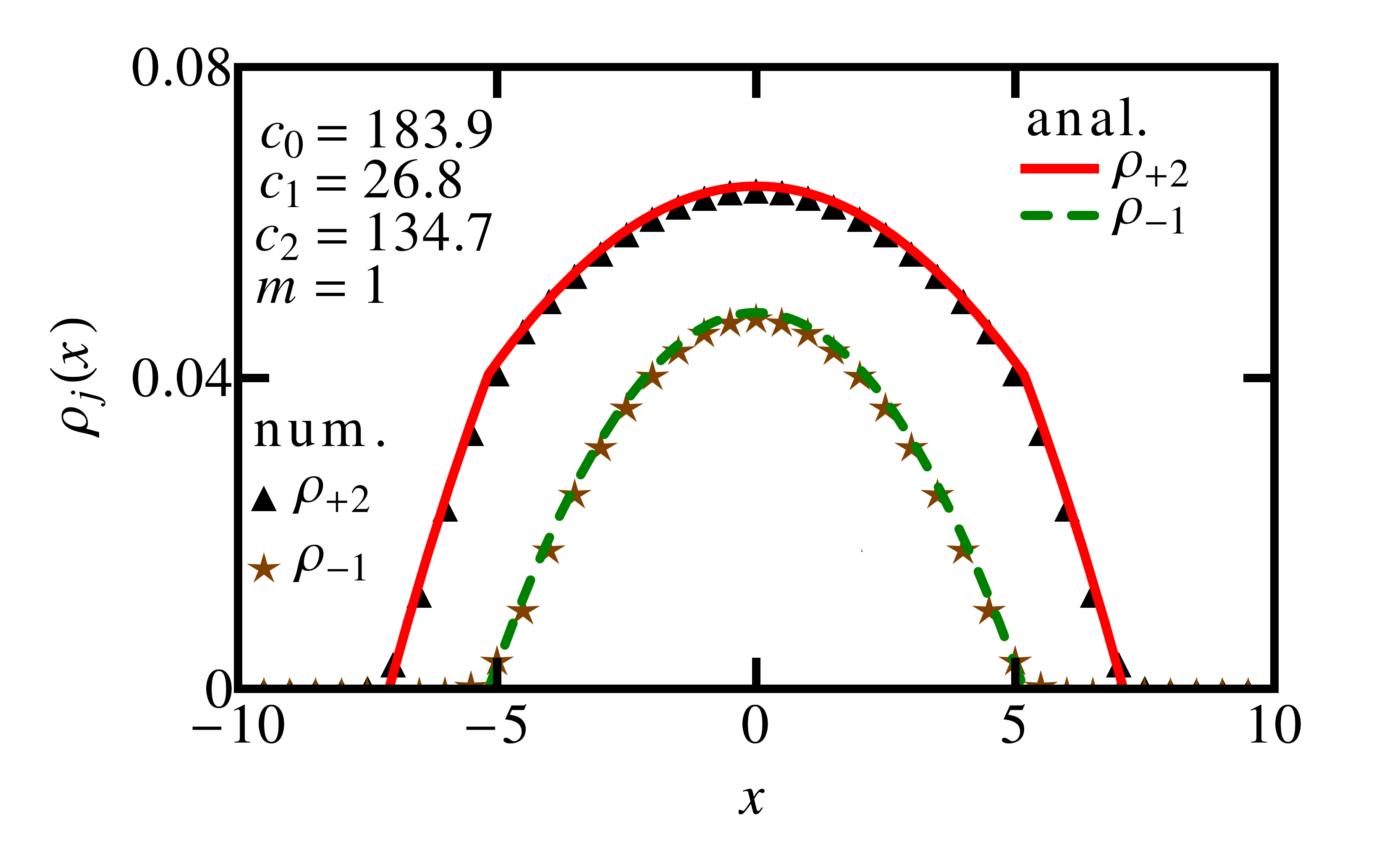}
\caption{(Color online) Analytic (anal.) and numerical (num.) densities of a spin-2 quasi-1D cyclic 
$^{23}$Na BEC.
The number of atoms, scattering lengths and oscillator lengths
are, respectively, $N = 10,000$, $a_0 = 34.9a_B$, $a_2 = 22.9a_B$, $a_4 = 64.5a_B$ \cite{Ciobanu}, $l_0 = 4.69\mu$m, $l_{yz} = 1.05\mu$m. The experimental value
of $a_2 (= 45.8a_B)$ has been modified to access the cyclic
phase of $^{23}$Na (using a Feshbach resonance) from its natural
anti-ferromagnetic phase.
}
\label{fig-4} \end{center}
\end{figure}
 
In this case, there are two degenerate ground states for all magnetization $m$ with non-zero 
component densities 
given by   Eqs. (\ref{dist-5}) and (\ref{dist-6}), respectively. The analytic models will 
be obtained in these two cases from the TFA to the GP equations (\ref{gp_s1})-(\ref{gp_s3}).

The former distribution  (\ref{dist-5}) involves only two non-zero components in the GP equation.
After neglecting the kinetic energy terms in the GP equations (\ref{gp_s1})-(\ref{gp_s3}),  the TFA   densities for the non-zero spin components ${m_f}=+2$ and ${m_f}=-1$ are 
  described by 
\begin{eqnarray}\label{tfx}
\mu_{+2}=X^2/2 +c_0 \rho + 2c_1(2\rho_{+2}-\rho_{-1}),\\
\mu_{-1}=X^2/2 +c_0 \rho - c_1(2\rho_{+2}-\rho_{-1}).\label{tfy}
\end{eqnarray}
For a non-zero magnetization $(2>m>0)$
the ${m_f}=+2$ component has a larger spatial 
extension   ($ l_{{\cal D}(+2)}$) than the ${m_f}=-1$ component 
with a smaller spatial extension ($\pm l_{{\cal D}(-1)}, l_{{\cal D}(+2)}>l_{{\cal D}(-1)}$). 
Following the procedure discussed for a 
spin-1   anti-ferromagnetic BEC in Sec. \ref{IIB}, one  obtains
%and $l_{{\cal D}(-1)}$ as
\begin{align}
l_{{\cal D}(+2)} &= l_{\cal D} {[(c_0+2 c_1 m)/{\cal C}_{I }]^{1/	(2+\cal D)}} \label{lp2_spin2_1d},\\
%l_{{\cal D}(+2)} &= \frac{[4 (c_0+2 c_1 m)]^{1/4}}{\pi ^{1/4}},\label{lp2_spin2_2d}\\
%l_{{\cal D}(+2)} &= \frac{[15(c_0+2 c_1 m)]^{1/5} }{(4\pi)^{1/5}}\label{lp2_spin2_3d}.
%\end{align}
%Similarly, $l_{{\cal D}(-1)}$ in 1D, 2D, and 3D, respectively, are
%\begin{align}
l_{{\cal D}(-1)} &= l_{\cal D} {[c_0(2 - m)/2{\cal C}_{I }]^{1/	(2+\cal D)}}.\label{lm1_spin2_1d}
%l_{{\cal D}(-1)} &= \frac{[2c_0 (2-m)]^{1/4}}{\pi^{1/4}},\label{lm1_spin2_2d}\\
%l_{{\cal D}(-1)} &= \frac{[15c_0(2 - m)]^{1/5} }{(8\pi)^{1/5}}\label{lm1_spin2_3d}.
\end{align}}
The normalized densities are given by 
\begin{align}\label{rho_cyclic2}
\rho_{-1}(X) &=   \frac{(l_{{\cal D}(-1)}^2-X^2)}{3 c_0}, \quad X{\le }l_{{\cal D}(-1)},\\
\rho_{+2}(X) &= \frac{3 c_0l_{{\cal D}(+2)}^2-2 (c_0-2c_1) l_{{\cal D}(-1)}^2-{\cal C}_{II} X^2
}{6 c_0 {\cal C}_{II}},
                  X{\le }l_{{\cal D}(-1)}
 \\ &= \frac{l_{{\cal D}(+2)}^2-X^2}{2{\cal C}_{II}}.\quad   l_{{\cal D}(-1)}{\le }X{\le }l_{{\cal D}(+2)},\label{rho_cyclic}
\end{align} 
Equations (\ref{rho_cyclic2})-(\ref{rho_cyclic}) together with extensions 
given by Eqs. (\ref{lp2_spin2_1d})-(\ref{lm1_spin2_1d}) are the analytic  densities in this case.
 The analytic   and numerical  densities for a quasi-1D spin-2 cyclic $^{23}$Na BEC
% for $c_0=183.9,c_1=26.8, c_2=134.7$
are shown in Fig. \ref{fig-4}. {For $m=0$, SMA becomes exact for the cyclic
phase of spin-2 condensate, as the $c_1$ and $c_2$-dependent terms in Eqs. 
(\ref{tfx})-(\ref{tfy}) vanish.} 

\begin{figure}[!t]
\begin{center}
\includegraphics[trim = 0mm 0mm 0cm 1cm, clip,width=1.1\linewidth,clip]{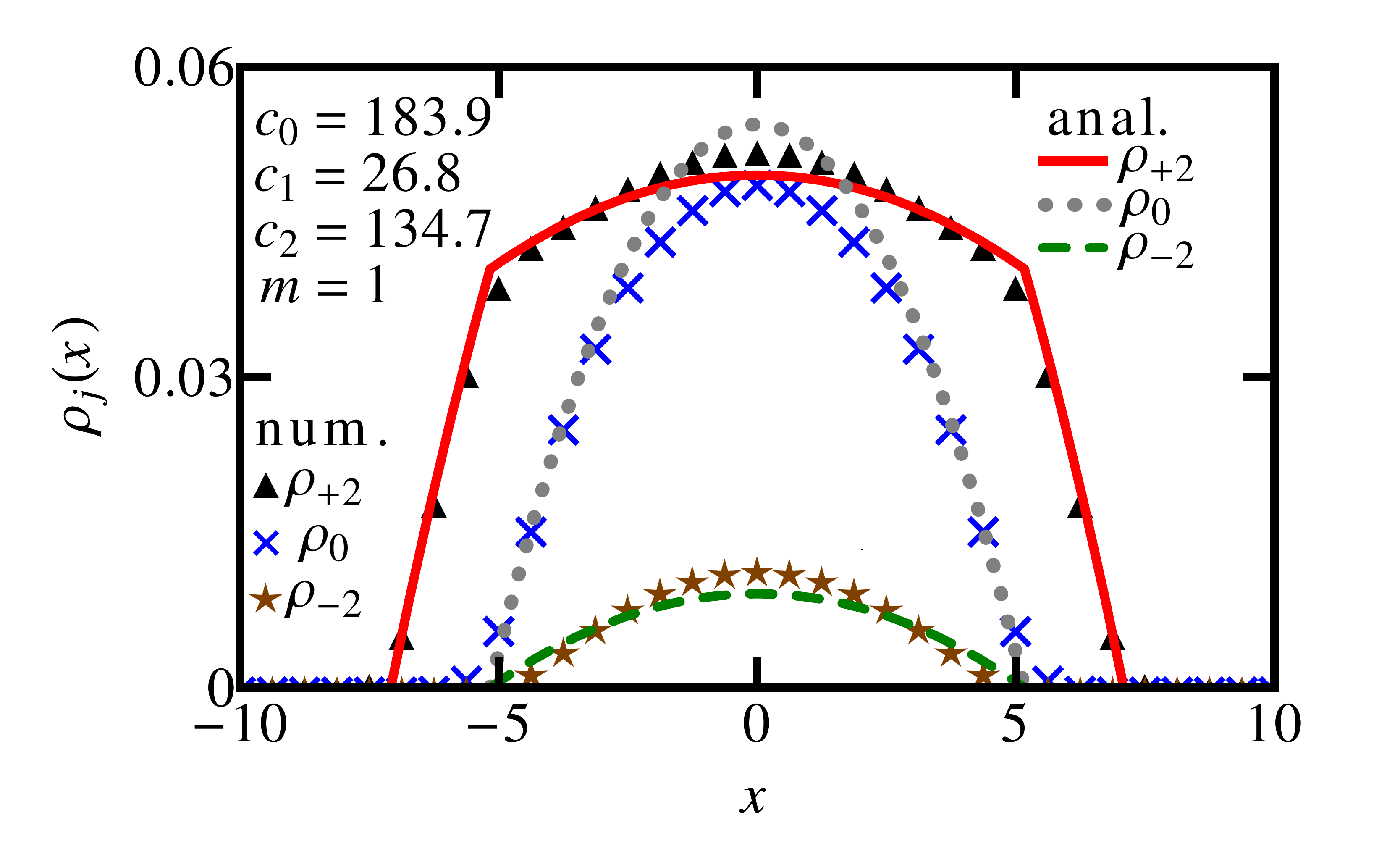}
\caption{(Color online) Analytic (anal.) and numerical (num.) densities of a spin-2 
quasi-1D $^{23}$Na cyclic BEC.
All parameters are the same as in Fig. \ref{fig-4}.}
\label{fig-5} \end{center}
\end{figure}

Comparing Eq. (\ref{tfasize}) and Eqs. (\ref{lp2_spin2_1d})-(\ref{lm1_spin2_1d}), 
the conditions for the validity of TFA in this case are
\begin{align}
\frac{N[4a_2+3a_4+2(a_4-a_2)m]}{7l_0}&>>1,\label{tfaspin2p2}\\
\frac{N[4a_2+3a_4(1-m/2)]}{7l_0}&>>1,\label{tfaspin2m1}
\end{align}
for $m_f=+2$ and $m_f=-1$ components, respectively. For $m=0$ the simple criterion
for the validity of TFA is $N(4a_2+3a_4)/(7l_0)>>1$, which is consistent
with fact that only spin-independent non-linearity ($c_0$ dependent term) contributes to 
the energy of the system.

Similarly, for the latter distribution (\ref{dist-6}), after neglecting the kinetic energy terms in the GP equations (\ref{gp_s1})-(\ref{gp_s3}),
 the  TFA densities  are given by
\begin{align}
 \mu_{\pm 2}  =&  
 X^2/2+c_0 \rho 
 \pm 4c_1 (\rho_{+2}-\rho_{-2}), \label{gps_cyc_1}\\
\mu_0  =&  
 X^2/2+c_0 \rho \label{gps_cyc_2}.
\end{align}
This set of equations for  $\rho_{\pm 2}$ and $\rho_0$ is overcomplete and does not determine the densities.  However, 
if we assume, consistent with Eq. (\ref{dist-6}),  that 
\begin{align}\label{z10}
\rho_0 = \frac{2 (2+m)\rho_{-2}}{2-m},
\end{align}
then we can solve Eqs. (\ref{gps_cyc_1}) for $\rho_{\pm 2}$ and 
  obtain $\rho_0$ from Eq. (\ref{z10}).
For $2>m>0$ the spatial extent ($l_{{\cal D}(+2)}$)
of  density $\rho_{+2}$ is larger 
 than  the spatial extent  ($l_{{ \cal D}(-2)}$)  of density $\rho_{-2}$.
Equations  (\ref{gps_cyc_1}) can  then be be solved  
to obtain
\begin{align}\label{rhop2spin2}
\rho_{+ 2}(X) = &\frac{4 c_1 (m-2) (X^2-\delta)+c_0 \kappa(6+m) }
{64 c_
0 c_1}, \quad X{ \le }l_{{ \cal D}(-2)}
\\
   = &  \frac{2\mu_{+2}-X^2}{2{\cal C}_{II}}, \quad 
l_{{\cal D}(+2)}{ \ge }X{ \ge }l_{{\cal D}(-2)},
\\%\end{align}
%\begin{align}
\rho_{- 2}(X) = &\frac{(m-2) \big[c_0 \kappa+4c_1 \{X^2- \delta\}\big]}{64 c_0 c_1}, \quad X{ \le }l_{ {\cal D}(-2)},\label{rhom2spin2}
\end{align}
where $\delta = \mu_{+2}+\mu_{-2}, \kappa = \mu_{+2}-\mu_{-2}.$
 with the chemical potentials  $\mu_{+2}$ and $\mu_{-2}$ given by 
\begin{align}
\mu_{+2}= \frac{l_{{\cal D}(+2)}^2}{2},\quad 
\mu_{-2}=\frac{(c_0-4 c_1) l_{{\cal D}(+2)}^2+8 c_1 l_{{\cal D}(-2)}^2}{2 {\cal C}_{II}},\label{z11}
\end{align}
where $l_{{\cal D}(+2)}$ and $l_{{\cal D}(-2)}$ are the same as $l_{{\cal D}(+2)}$ and $l_{{\cal D}(-1)}$
of Eqs. (\ref{lp2_spin2_1d})-(\ref{lm1_spin2_1d}), 
respectively.
After substituting the expressions for chemical potentials 
$\mu_{\pm 2}$ given by Eqs. (\ref{z11}) in Eqs.  
 (\ref{z10})-(\ref{rhom2spin2}), we obtain the final   densities $\rho_{\pm 2}$ and $\rho_0$ as   
%\begin{widetext}
\begin{eqnarray}
\rho_{+2}(X) &=& \frac{
 c_0 \left[8 l_{{\cal D}(+2)}^2 - l_{{\cal D}(-2)}^2 (6 + m) + (-2 + m) X^2\right]}{16 c_0 {\cal C}_{II}}\nonumber
  \\
&+& \frac{4 c_1 (2 - m) (l_{{\cal D}(-2)}^2 - X^2) }{16 c_0 {\cal C}_{II}},\quad X {\le } l_{{\cal D}(-2)},
\end{eqnarray}
%\end{widetext}
\begin{eqnarray}
\rho_{+2}(X) &=& \frac{l_{{\cal D}(+2)}^2-X^2}{2 {\cal C}_{II}},\quad l_{{\cal D}(+2)}{\ge }X{\ge }l_{{\cal D}(-2)},
\\
\rho_{0}(X)&=&\frac{2 (2+m)\rho_{-2}}{2-m}, \quad X {\le } l_{{\cal D}(-2)},
\end{eqnarray}
\begin{eqnarray}
\rho_{-2}(X)&=& \frac{(2-m) \left(l_{{\cal D}(-2)}^2-X^2\right)}{16 c_0}, \quad X {\le } l_{{\cal D}(-2)}\\
\rho_{0}(X)&=& \rho_{-2} =0,\quad l_{{\cal D}(+2)}{\ge }X{\ge }l_{{\cal D}(-2)}.
\end{eqnarray}
 
The analytic  and numerical
densities in this case  for a spin-2 quasi-1D $^{23}$Na cyclic BEC are shown in Fig. \ref{fig-5}.
The criteria for the validity of TFA in this case are again given by Eqs. (\ref{tfaspin2p2})-(\ref{tfaspin2m1})
for $m_f = +2$ and $m_f = -2$ components, receptively.
Thus, for a spin-2 $^{23}$Na cyclic  BEC, there are two distinct degenerate
ground states
as are shown in Figs. \ref{fig-4} and \ref{fig-5}. The Hamiltonian of the spinor 
BEC is time-reversal invariant, yet the degenerate states shown in  Figs. \ref{fig-4} and \ref{fig-5} break time reversal symmetry. Time-reversal symmetry-breaking states 
in spinor BECs were previously studied \cite{GA-2}. {In cyclic phase too the additional potential
energy cost rules out the possibility of asymmetric phase-separated profiles as ground states}.

\section{Concluding Remarks}
\label{V}

The mean-field GP equation for a spin-1 and spin-2 spinor BEC involve 
three- and five-component complex wave function. Some simplification emerges for the ground-state wave function  of a spinor BEC. 
For an anti-ferromagnetic or cyclic BEC with a non-zero magnetization,   some of the spin-component wave functions 
become zero, thus reducing the original GP equation with three or five components 
to a system of coupled equations with only two or three components, which we call a reduced GP equation. 
For a ferromagnetic BEC with a non-zero magnetization the densities of different spin components 
for 
the ground-state wave function are found to be multiples of each other. This allows to solve the density 
according to a single GP equation, which we call the decoupled-mode  (DM) equation,   and calculate the densities of different spin components as multiples 
of a single DM density. These reduced GP and DM equations are valid in all spatial 
dimensions. 
Here we suggest simple analytic models for the ground-state densities of 
a  
spinor BEC obtained by applying Thomas-Fermi approximation to the DM and reduced GP equations.  These 
analytic results for densities are found to be in good agreement with those obtained from the 
numerical solution of the full GP equation for  ferromagnetic, anti-ferromagnetic, and cyclic   spin-1 and spin-2 spinor BECs. 
Although, we considered in this paper nearly-overlapping configurations of the spinor
components, {the presence of Zeeman energy and spin-orbit coupling in the
Hamiltonian can lead to asymmetric phase-separated configurations \cite{sadhan,GA-2} {as ground states}.}
%there could be for certain parameter domains a complete phase separation 
%\cite{sadhan}.
%among the components.  
An investigation leading to the analytic densities of the phase-separated solutions 
would be an interesting future work.

\begin{acknowledgements}

This work is financed by the Funda\c c\~ao de Amparo \`a Pesquisa do Estado de 
S\~ao Paulo (Brazil) under Contract Nos. 2013/07213-0, 2012/00451-0 and also by 
the Conselho Nacional de Desenvolvimento Cient\'ifico e Tecnol\'ogico (Brazil).
\end{acknowledgements}

\section{Appendix A}

\subsection{Ferromagnetic spin-1 BEC}
For the ground state of  a  spin-1 ferromagnetic BEC ($c_1<0$), the coefficients $\alpha_j,$ can  be obtained from a minimization of the energy
\begin{align}
 E =&  \frac{N}{2}\int \Big[\sum_{j={-1}}^1
| \phi_j'|^2  
 +  {x^2\rho+c_0\rho^2 + c_1|\mathbf F|^2}\Big]d{\bf x},
  \label{energy_spin1}
\end{align}
where prime denotes $x$ derivative.
%For $c_1<0$, energy can be minimized by maximizing $\int |\mathbf F|^2 dx$.
Assuming that component wave functions are given by the DM {\em ansatz} (\ref{sma}), to 
minimize energy $E$ we need to  
  maximize the positive integral  $ \int |\mathbf F|^2 d{\bf x} $ 
\begin{align}
 \int |\mathbf F|^2d{\bf x} &=\big[2|(\alpha_{+1}^*\alpha_0+\alpha_0^*\alpha_{-1})|^2+m^2\big] 
                     {\cal I}, \\
{\cal I}&=\int\phi_{\rm DM}^4(x)d{\bf x}.
\end{align}
  Now, writing $\alpha_j = |\alpha_j|e^{i\theta_j}$,
we get
\begin{eqnarray}\label{t1}
\int |\mathbf F|^2d{\bf x} &=& \big[2||\alpha_{+1}||\alpha_0|+|\alpha_0||\alpha_{-1}|e^{i(\theta_{+1}+\theta_{-1}-2\theta_0)}|^2
                   \nonumber\\&& +m^2\big]{\cal I}
\label{fspin1a}.
\end{eqnarray}
To maximize  integral (\ref{t1}),   we take $\exp (\theta_{+1}+\theta_{-1}-2\theta_0)=1$ and obtain 
\begin{eqnarray}
{ \int} |\mathbf F|^2d{\bf x} = \big[2(|\alpha_{+1}||\alpha_0|+|\alpha_0||\alpha_{-1}|)^2
                   +m^2\big]{\cal I} 
\label{fspin1b}.
\end{eqnarray}
For a fixed magnetization $m$ and DM function $\phi_{\mathrm{DM}}$, the 
maximization of $\int |\mathbf F|^2 d{\bf x}$ corresponds to finding 
the stationary points
of the following ``Lagrange'' function
\begin{align}\label{lagr}
&L_1(|\alpha_{j}|,\lambda_1, \lambda_2) = 2(|\alpha_{+1}| |\alpha_{0}|  +|\alpha_{0}|  |\alpha_{-1}| )^2 \nonumber\\
&+\lambda_1(1-\sum_{j}|\alpha_j|^2)
+\lambda_2(m-|\alpha_{+1}|^2+|\alpha_{-1}|^2). 
\end{align}
Here $\lambda_1$ and  $\lambda_2$  are Lagrangian multipliers to fix
the normalization  and  magnetization
to $1$ and  $m$, respectively.
The stationary points of $L_1$ are determined by the following Lagrange equations
\begin{eqnarray}
\frac{\partial L_1}{\partial |\alpha_j|}=  0,\quad
 \frac{\partial L_1}{\partial \lambda_1} = 0, \quad \frac{\partial L_1}{\partial \lambda_2}=0.
\end{eqnarray}
with solution (\ref{dist-1}) together with
{$\lambda_1= 2,  \lambda_{2} = -2m.$}

\subsection{Anti-ferromagnetic spin-1 BEC}

In case of an anti-ferromagnetic BEC ($c_1>0$), $\int |\mathbf F|^2 d{\bf x} = \int (F_{+}F_{-}+F_z^2)d{\bf x}$  is minimized
by making $\phi_0(x) =0$ for any $m\ne0$ and  the densities 
satisfy Eq. (\ref{norm-1}). 
If we further assume the DM {\em ansatz} (\ref{sma}), the coefficients $\alpha_j$ can be obtained 
from a minimization of (\ref{fspin1a}). For this, we take $\exp (\theta_{+1}+\theta_{-1}
-2\theta_0)=-1$ and obtain
\begin{eqnarray}
{ \int} |\mathbf F|^2d{\bf x} = \big[2(|\alpha_{+1}||\alpha_0|-|\alpha_0||\alpha_{-1}|)^2
                   +m^2\big]{\cal I} 
\label{fspin1v}.
\end{eqnarray}
Following the procedure discussed for a ferromagnetic BEC, one can minimize 
$\int |\mathbf F|^2d{\bf x}$ under the twin constraints of fixed norm and magnetization
and, in agreement with Eq. (\ref{norm-1}), obtain 
\begin{equation}
|\alpha_{\pm 1}| = \sqrt{\frac{1\pm m}{2}}, \quad \alpha_{0} = 0.
\end{equation}

\section{Appendix B}

\subsection{Ferromagnetic spin-2 BEC}

For a  spin-2 ferromagnetic BEC ($c_1<0, c_2>20c_1$), the energy is given by
\begin{align}
 E =&  \frac{N}{2}\int \Big[\sum_{j={-2}}^2
|\phi_j'|^2  
 +  {x^2\rho+c_0\rho^2 + c_1|\mathbf F|^2 + c_2|\mathbf \Theta|^2}\Big]d{\bf x}.\nonumber
 % \label{energy}
\end{align}
For a ferromagnetic BEC, the energy minimization 
 corresponds to  a maximization of the 
$c_1$-dependent term $\int |\mathbf F|^2 d{\bf x}$. We find that  this  automatically minimizes the $c_2$-dependent 
term $\int |\Theta|^2 d{\bf x}$ to zero. 
Assuming   the DM {\em ansatz} (\ref{comp-2})  we seek the 
coefficients $\alpha_j$ which maximize
$ \int |\mathbf F|^2 d{\bf x} $. 
Following the procedure for ferromagnetic
spin-1 BEC, we can write
\begin{align}\label{eqz2}
\int &|\mathbf F|^2 d{\bf x} = \Big[\Big|2\big\{|\alpha_{+2}|| \alpha_{+1}|  +|\alpha_{-2}|  |\alpha_{-1}|\nonumber \\
 &e^{i(\theta_{-2}-\theta_{-1}-\theta_{1}+\theta_2)} \big\} +\sqrt{6}e^{i(\theta_{0}-2\theta_{1}+\theta_2)} \nonumber \\
&\big\{  |\alpha_{+1}||\alpha_0| + |\alpha_0||\alpha_{-1}|e^{i(\theta_{-1}-2\theta_{0}+\theta_1)} \big\}\Big|^2+m^2\Big] {\cal I}.
\end{align} 
To maximize (\ref{eqz2}) we take all exponential factors in this equation to be $+1$ and obtain
\begin{align}%\label{eqz}
\int |\mathbf F|^2 d{\bf x}&= \Big[\Big\{2(|\alpha_{+2}| |\alpha_{+1}|  +|\alpha_{-2}|  |\alpha_{-1}| )
 \nonumber \\
&+
\sqrt{6}(  |\alpha_{+1}||\alpha_0|  
+ |\alpha_0||\alpha_{-1}| )\Big\}^2+m^2\Big]{\cal I}.
\end{align}
For a fixed  $m$ ($2>m>0$) and $\cal I$, the maximization of $\int |\mathbf F|^2 d{\bf x}$ 
corresponds to finding the stationary points of the   following Lagrange function
\begin{align}
L_2(|\alpha_{j}|,\lambda_1, \lambda_2) &= 2(|\alpha_{+2}| |\alpha_{+1}|  +|\alpha_{-2}|  |\alpha_{-1}| )+
\sqrt{6}|\alpha_0|  \nonumber \\
&(  |\alpha_{+1}|+ |\alpha_{-1}| ) +\lambda_1(1-\sum_{j}|\alpha_j|^2) +\lambda_2 \nonumber\\
&(m\nonumber-2|\alpha_{+2}|^2
-|\alpha_{+1}|^2+|\alpha_{-1}|^2+2|\alpha_{-2}|^2).
\end{align}
Here $\lambda_1$ and  $\lambda_2$ have the same meaning as in  Eq. (\ref{lagr}).
The stationary point which maximizes $\int|\mathbf F|^2d{\bf x}$  thus yields Eqs. (\ref{comp-3})-(\ref{comp-4}) together with
\begin{equation}
\lambda_1=\frac{4  }{ \sqrt{(4-m^2)}  },\quad
\lambda_2=-\frac{m   }{ \sqrt{(4-m^2)}  } .
\end{equation} 
Using Eqs. (\ref{comp-3})-(\ref{comp-4}), we find that 
these $\alpha_j$'s also
minimize  $\int|\Theta|^2d{\bf x}$ to 0 which guarantees that 
the state so obtained is the ground state.

\subsection{Anti-ferromagnetic spin-2 BEC}
Similarly in anti-ferromagnetic subdomain ($c_2<0$ and $c_2<20c_1$), the 
energy minimization corresponds to a maximization of the $c_2$-dependent
term $\int |\Theta|^2 d{\bf x}$. Assuming DM {\em ansatz} (\ref{norm-1}) and  
$\alpha_j=|\alpha_j|\exp(i\theta_j)$ we have   
\begin{eqnarray}\label{eqt}
 \int |\Theta|^2 d{\bf x}&=&\big|2|\alpha_{+2}| |\alpha_{-2}|  -2|\alpha_{+1}|  |\alpha_{-1}|e^{i(\theta_{1}+\theta_{-1}-\theta_{2}-\theta_{-2})} 
\nonumber\\&&+|\alpha_0|^2e^{i(2\theta_{0}-\theta_{2}-\theta_{-2})}\big|^2 \cal I.
\end{eqnarray}
To maximize integral  (\ref{eqt}) we take the first exponential to be $-1$ and the second exponential 
to be $+1$. 
%Thus, in the case of a anti-ferromagnetic BEC, the phases must satisfy
%\begin{eqnarray}
%\theta_{1}+\theta_{-1}-\theta_{2}-\theta_{-2} &= (2n+1)\pi, \\
%2\theta_{0}-\theta_{2}-\theta_{-2} &= 2n\pi, 
%\end{eqnarray}
%where $n$ is an integer.
 For a fixed  $m$ ($2>m>0$) and $\cal I$,  the 
maximization of $\int |\Theta|^2 d{\bf x}$ 
corresponds to finding the stationary points of the following Lagrange function
\begin{eqnarray}
L_{\theta}(|\alpha_{j}|,\lambda_1, \lambda_2) &=& (2|\alpha_{+2}| |\alpha_{-2}|  +2|\alpha_{+1}|  |\alpha_{-1}|+|\alpha_0|^2)\nonumber
\\&&+\lambda_1(1-\sum_{j}|\alpha_j|^2) +\lambda_2(m-2|\alpha_{+2}|^2  \nonumber \\
&&-|\alpha_{+1}|^2+|\alpha_{-1}|^2+2|\alpha_{-2}|^2).
\end{eqnarray}
 The stationary point, which maximizes $\int|\Theta|^2d{\bf x}$,    yields 
\begin{eqnarray}\label{kk}
|\alpha_{\pm 2}| &=& \frac{\sqrt{2\pm m}}{2},\quad \alpha_{\pm 1} = \alpha_{0} = 0, \\
\lambda_1&=&\frac{2  }{ \sqrt{(4-m^2)}  },\quad
\lambda_2=\frac{-m   }{ 2\sqrt{(4-m^2)}  } . \label{ll}
\end{eqnarray}
Using Eqs. (\ref{kk})-(\ref{ll}), we find that $\int |\mathbf F|^2d{\bf x}$ has the minimum value 
$m^2\cal I$ which guarantees that
the state so obtained is the ground state.

{\subsection{Cyclic spin-2 BEC}}

For a spin-2 cyclic BEC $c_1>0$ and $c_2>0$, energy minimization involves minimization
of both $\int |\mathbf F|^2d{\bf x}$ and $\int |\Theta|^2d{\bf x}$  to their respective 
minimum values $m^2{\cal I}$ and 0, respectively.
From equations (\ref{eqz2}) and (\ref{eqt}), one can see that for $0<m<2$, consistent 
with Eqs. (\ref{dist-5})-(\ref{dist-6}),
there are only two 
possibilities for the ground states:
\begin{equation} 
(i)|\alpha_{\pm 2}|= \frac{{2\pm m}}{{4}},\quad |\alpha_0|=\frac{\sqrt{4-m^2}}
{\sqrt{8}},\quad \alpha_{\pm 1}=0,\nonumber 
\end{equation}
 provided that 
$\exp(2\theta_0-\theta_{+2}-\theta_{-2})=-1$, and 
\begin{equation}
(i) |\alpha_{+2}| = \frac{\sqrt{1+m}}{\sqrt{3}}  , \quad |\alpha_{-1}|=\frac{\sqrt{2-m}}{\sqrt{3}}, \quad \alpha_{\pm 1}=\alpha_0=0.\nonumber 
\end{equation}

\end{document}